\documentclass[aps,prd,twocolumn,preprintnumbers]{revtex4-1}
\usepackage{graphicx}
\usepackage{hyperref}
\usepackage{amsmath}
\usepackage{diagbox}
\usepackage{slashed}
\usepackage{epstopdf}

\newcommand{\sign}[1]{\text{sign}\left( #1 \right)}
\newcommand{\hc}{\text{h.c.}}

\newcommand{\sD}{s\hspace{-0.06cm}D}
\newcommand{\fD}{f\hspace{-0.09cm}D}
\newcommand{\vD}{v\hspace{-0.08cm}D}
\newcommand{\rD}{r\hspace{-0.08cm}D}
\newcommand{\tD}{t\hspace{-0.08cm}D}
\newcommand{\sB}{s\hspace{-0.06cm}B}
\newcommand{\fB}{f\hspace{-0.09cm}B}
\newcommand{\vB}{v\hspace{-0.08cm}B}
\newcommand{\rB}{r\hspace{-0.08cm}B}
\newcommand{\tB}{t\hspace{-0.08cm}B}

\usepackage{color}

\begin{document}

\preprint{PITT-PACC-1311}
\title{A New Method for the Spin Determination of Dark Matter}

\date{\today}

\author{Neil D. Christensen}
\author{Daniel Salmon}

\affiliation{Pittsburgh Particle physics, Astrophysics and Cosmology Center (PITT PACC), Department of Physics $\&$ Astronomy, University of Pittsburgh, 3941 O'Hara St., Pittsburgh, PA 15260, USA}

\begin{abstract}
We construct a new kinematical variable that is able to fully reconstruct the absolute value, and partially reconstruct the sign, of the angular distribution in the center of momentum system of a decaying particle in certain cases where the center of momentum system is only known up to a two-fold ambiguity.  After making contact with Drell-Yan production at the Large Hadron Collider, we apply this method to the pair-production of dark matter in association with two charged leptons at the International Linear Collider and show that for a small intermediate width, perfect agreement is found with the true angular distribution in the absence of initial state radiation.  In the presence of initial state radiation, we find that the modification to the angular distributions is small for most angles and that different spin combination classes should still be distinguishable.  This enables us to determine the spin of the mother particle and the dark matter particle in certain cases.
\end{abstract}

\maketitle

The existence of dark matter has been well established through a combination of galactic rotation curves \cite{Zwicky:1933gu,Rubin:1970zza,Rubin:1980zd,Bosma:1981zz}, weak and strong gravitational lensing \cite{Refregier:2003ct,Tyson:1998vp}, Big Bang nucleosynthesis \cite{Olive:1999ij}, the cosmic microwave background \cite{Komatsu:2010fb} and the bullet cluster \cite{Clowe:2006eq}. From these observations, we know that dark matter is electrically neutral, non-baryonic and composes roughly 83\% of the matter and 23\% of the energy of the universe.  However, these observations do not tell us the detailed properties of dark matter such as its mass, spin and how it interacts with visible matter.  For that, we need to observe a dark matter particle (DMP) in the laboratory.

Because the Standard Model (SM) of particle physics does not contain dark-matter (among other things) it is a low-energy effective theory that fits inside a larger, more complete theory.  Two prominent examples of these theories are the minimal supersymmetric extension of the SM (MSSM) and the universal extra-dimension (UED) model.  In the present context, one of the most important features of these models is the presence of a new parity symmetry with the consequence that the lightest parity-odd particle (LPP) is stable and (if neutral) a dark-matter candidate \cite{Goldberg:1983nd,Ellis:1983ew,Cheng:2002ej,Servant:2002aq}.  In these theories, the LPP is a weakly interacting massive particle (WIMP) and, so, can be pair produced at particle colliders, such as the Large Hadron Collider (LHC) and the International Linear Collider (ILC).

\begin{figure}
\includegraphics[scale=.5]{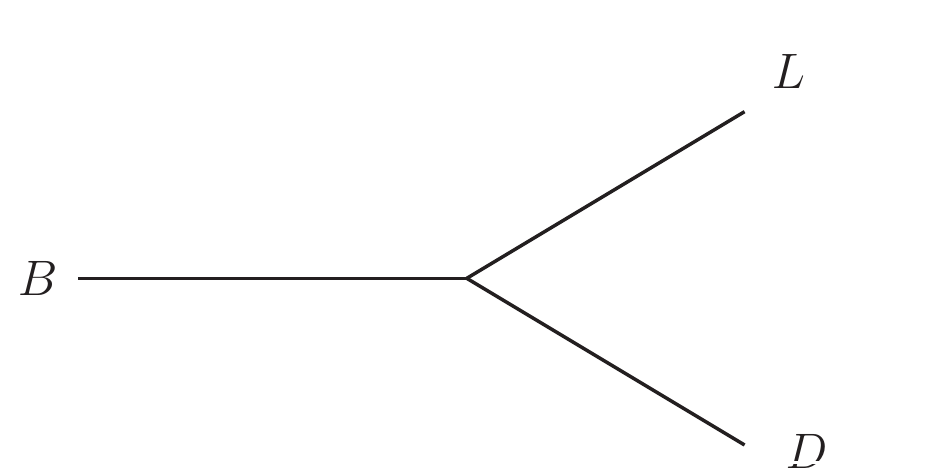}
\caption{\label{fig:bldsystem} Decay of $B$ into $L$ and $D$ where $L$ is an observed SM-particle, $D$ is the missing dark-matter particle and $B$ is the parent of this decay.}
\end{figure}

To determine the spin of a DMP at a collider, ideally, we would like to boost into the center of momentum (CM) frame of its parent particle and histogram the angle of its decay with respect to the boost direction (see Figure \ref{fig:bldsystem}).  We will call this the CM angular distribution, where $\theta_{LB}$ is the angle of the decay product $L$ with respect to the boost direction in the $B$ CM system.  If the width of the parent particle is narrow, this distribution will correspond with linear combinations of squares of the Wigner $d^j_{m,m'}$-functions where $j$ and $m$ correspond with the spin and spin-component along the boost direction of the parent particle $B$ and $m'$ corresponds with the difference of the helicities of the final state particles $L$ and $D$ (see Appendix \ref{sec:Wigner d-functions} for a brief discussion.)  
The challenge for dark-matter particles is that they do not interact with particle detectors and are, thus, not measured.  Therefore, since we do not know their momentum, we often can not reconstruct the CM system.  

In this paper, we introduce a new kinematical variable that is able to fully reconstruct the absolute value of the CM angular distribution unambiguously and its sign up to a two-fold ambiguity even in some cases where the CM system is not known.  This method is a generalization of that used to reconstruct the spin of a new charged resonance in Drell-Yan processes at the LHC \cite{Chiang:2011kq}.  Our result is the following.  The absolute value of the cosine of this angle is given by
\begin{equation}
|\cos{\theta _{LB}}| = \sqrt{1-\frac{4M_B^2p_{\widetilde{T}}^2}{\lambda \left( M_B^2,M_L^2,M_D^2 \right)}} \label{eq:cosmagfinal}\ ,
\end{equation}
where $M_B$, $M_L$ and $M_D$ are the masses of the parent particle ($B$), the observed particle ($L$) and the DMP ($D$), $p_{\widetilde{T}}$ is the component of the observed momentum of the visible particle ($L$) that is transverse to the momentum of $B$ in the lab-frame, and 
\begin{equation}
\lambda \left( x,y,z \right)=x^2+y^2+z^2-2xy-2xz-2yz\ \label{eq:lambda}.
\end{equation}
The two possible signs are given by
\begin{equation}
\cos{{\theta}_{LB\mathcal{L}}} =-|\cos{{\theta}_{LB}}| \label{eq:largesolution}
\end{equation}
and
\begin{equation}
\cos{{\theta}_{LB\mathcal{S}}}=\sign{E_{L}-\frac{M_B^2+M_L^2-M_D^2}{2M_B}} |\cos{{\theta}_{LB}}| \label{eq:smallsolution}\ ,
\end{equation}
which have been labeled by $\mathcal{L}$ and $\mathcal{S}$, to be explained below and $E_L$ is the energy of the observed particle ($L$) in the lab frame.

Furthermore, $p_{\widetilde{T}}$ can be expressed purely in terms of known quantities and the energy of $D$ in the lab frame as
\begin{equation}
p^2_{\widetilde{T}}=\frac{\left(E_L^2-M_L^2\right)\left(E_D^2-M_D^2\right)-\xi^{2}}{\left(E_L+E_D\right)^2-M_B^2}\label{eq:pTED1}\ ,
\end{equation}
where
\begin{equation}
\xi=\frac{1}{2}\left(M_L^2+M_D^2-M_B^2+2E_LE_D\right)\label{eq:pTED2}\ .
\end{equation}


The requirements for this method are that:  the masses of $B$, $L$ and $D$ must be known; the full momentum of $L$ must be known; the width of $B$ must be narrow; $B$, $L$ and $D$ must all be on-shell; and either $p_{\widetilde{T}}$ or $E_D$ must be known.  Under these circumstances, even if the B CM system can not be reconstructed, the CM angular distribution can be calculated, up to the sign ambiguity outlined above.  In this paper, we will describe two scenarios where these requirements are satisfied.  The first is in the discovery of a new resonance in charged Drell-Yan production of a charged lepton and a neutrino at the LHC.  We will summarize this scenario and refer to \cite{Chiang:2011kq} for further details.  The second is in the Antler production of two charged leptons and two dark matter particles at the ILC, which we will describe in detail in this paper.

Before moving on, we give a brief summary of other methods to measure the spin of dark matter.  An analysis of the spin-correlation in various cascade decay chains has shown that in many cases the resulting distributions were sufficient to determine the spin \cite{Barr:2004ze,Datta:2005zs,Smillie:2005ar,Wang:2006hk,Burns:2008cp,Gedalia:2009ym,Bai:2010hd}.  It has also been found that in certain cases the production cross-section varies with spin \cite{Datta:2005vx,Ginzburg:2010tu}.  Additionally, the shapes of some other distributions have a dependence on the spin \cite{Meade:2006dw,Buckley:2007th,Cho:2008tj,Buckley:2008eb,Ehrenfeld:2009rt,Chen:2011cya,Ginzburg:2012qa,Edelhauser:2012xb,Guadagnoli:2013xia}.  In addition to these methods, our method has the benefit of reconstructing the actual CM angular distribution even when the CM system can not be reconstructed, in many cases.

The rest of this paper is organized as follows. In Section~\ref{sec:derivation}, we derive these kinematical variables in detail.  In Section~\ref{sec:dylhc}, we summarize the application of these methods to charged Drell-Yan production and consider the effects of a finite width.  In Section~\ref{sec:antlers}, we describe how the application of our kinematical variables to antler processes at the ILC can be used to determine the spin of dark matter.  In Section~\ref{sec:conclusion}, we conclude.  


\section{\label{sec:derivation}Derivation}
In this section, we summarize the derivation of Eqs.~\eqref{eq:cosmagfinal} through \eqref{eq:pTED2}.   We consider an on-shell particle $B$ which decays to the particles $L$ and $D$, both of which are on-shell (see Fig.~\ref{fig:bldsystem}).  We assume the masses of these particles are known and are $M_B$, $M_L$ and $M_D$, respectively.  
We boost into the $B$ CM frame and calculate the cosine of the angle of $L$ with respect to the boost direction, which is given by
\begin{equation}
\cos{\theta_{LB}}=\frac{\vec{p}_{B} \cdot \vec{p}_{LCM}}{|\vec{p}_{B}||\vec{p}_{LCM}|}\ ,
\end{equation}
where the subscript $CM$ refers to the $B$ CM frame.  For convenience, and without loss of generality, we take the $z$-direction to be in the same direction as $\vec{p}_B$, the momentum of $B$.  Therefore, this can be rewritten as
\begin{equation}
\cos{\theta_{LB}}=\sign{p_{LzCM}}\frac{|p_{LzCM}|}{|\vec{p}_{LCM}|}\label{eq:cosdef}\ ,
\end{equation}
where $p_{Lz}$ refers to the z-component of the $L$ momentum.

\subsection{\boldmath{$|\cos{\theta_{LB}}|$}}

We will first consider the magnitude of Eq. \eqref{eq:cosdef}, which is given by
\begin{equation}
|\cos{\theta_{LB}}|=\frac{|p_{LzCM}|}{|\vec{p}_{LCM}|} \label{eq:cosmag}\ .
\end{equation}
For a massive $L$ we have
\begin{eqnarray}
|p_{LzCM}|&=&\sqrt{E_{LCM}^2-M_L^2-p_{\widetilde{T}}^2}\ ,\\
|\vec{p}_{LCM}|&=&\sqrt{E_{LCM}^2-M_L^2}\ ,
\end{eqnarray}
where $p_{\widetilde{T}}$ indicates the component of $\vec{p}_L$ that is transverse to $\vec{p}_{B}$.  We note that $p_{\widetilde{T}}$ is invariant under a boost into the $B$ CM frame, and is therefore the same in the lab frame as in the $B$ CM frame. Simplifying, we have\begin{equation}
|\cos{\theta_{LB}}|=\sqrt{1-\frac{p_{\widetilde{T}}^2}{E_{LCM}^2-M_L^2}} \label{eq:ptmaxancillary}\ .
\end{equation}
Conservation of momentum in the $B$ CM frame leads to
\begin{equation}
E_{LCM}=\frac{M_B^2+M_L^2-M_D^2}{2M_B} \label{eq:elcms}\ .
\end{equation}
Substituting this gives
\begin{equation}
|\cos{\theta}_{LB}|=\sqrt{1-\frac{4M_B^2p_{\widetilde{T}}^2}{\lambda{\left(M_B^2,M_L^2,M_D^2\right)}}}\ ,
\end{equation}
which completes the derivation of Eq.~\eqref{eq:cosmagfinal}.
We note that this is expressed in terms of lab-frame quantities and invariants only.

\subsection{\boldmath{$\sign{p_{LzCM}}$}}
We now turn to the sign of $p_{LzCM}$. We begin by boosting $p_L$ into the $B$ CM frame giving
\begin{equation}
\sign{p_{LzCM}}=\sign{E_Bp_{Lz}-E_L p_{Bz}}\ \label{eq:sign(plzcms)}.
\end{equation}
Using the solution for $p_{Dz}$ from Appendix~\ref{sec:pdz} and $p_{Bz}=p_{Lz}+p_{Dz}$, we have
\begin{equation}
p_{Bz}=\frac{M_BE_{LCM}}{M_L^2+p_{\widetilde{T}}^2}\left(p_{Lz}\pm E_L\zeta\right) \label{eq:pbz}\ ,
\end{equation}
where
\begin{equation}
\zeta = \sqrt{1-\frac{M_L^2+p^2_{\widetilde{T}}}{E^2_{LCM}}}\label{eq:sigma}\ ,
\end{equation}
We have two sign possibilities given by the $\pm$.  When this sign matches the sign of $p_{Lz}$, we will call this the ``large" solution and use the symbol $\mathcal{L}$ to denote it while if the sign is opposite that of $p_{Lz}$, we will call this the ``small" solution and use the symbol $\mathcal{S}$ to denote it.  

If $p_{Lz}<0$, since $p_{Bz}>0$ (by definition of the $z$-direction), we must have the $+$ sign and be in the small solution.  Therefore according to Eq.~\eqref{eq:sign(plzcms)}
\begin{equation}
\sign{p_{LzCM}}_{\mathcal{S}<}=-1\label{eq:sign pLzCM small minus}\ ,
\end{equation}
where the $<$ refers to the fact that this is for $p_{Lz}<0$.  

On the other hand, if $p_{Lz}>0$, Eq.~\eqref{eq:sign(plzcms)} is the sign of the difference of two positive terms.  This sign is the same as the sign of the difference of these two positive terms squared giving
\begin{equation}
\sign{p_{LzCM}}_>=\sign{E_B^2p_{Lz}^2-E_L^2 p_{Bz}^2}\ \label{eq:sign(plzcms>)}.
\end{equation}
Using $E_B^2=M_B^2+p_{Bz}^2$ and $E_L^2-p_{Lz}^2=M_L^2+p_{\widetilde{T}}^2$, gives us
\begin{equation}
\sign{p_{LzCM}}_> = \sign{M_B^2p^2_{Lz}-\left( M_L^2+p^2_{\widetilde{T}}\right)p^2_{Bz}}\ .\label{eq:sign(plzcms)2}
\end{equation}
By substituting Eq. \eqref{eq:pbz} for $p_{Bz}$ this can be expanded and simplified to the form 
\begin{equation}
\sign{p_{LzCM}}_> = \sign{-\left(p^2_{Lz}+E_L^2\right)\zeta^2 \mp 2p_{Lz}E_L\zeta}\ \label{eq:signplzcmsint1}.
\end{equation}
The large solution corresponds to the top sign choice, which is a minus sign, giving us
\begin{equation}
\sign{p_{LzCM}}_{\mathcal{L}}=-1\ ,
\end{equation}
which completes the derivation of Eq.~\eqref{eq:largesolution}.
However, it still remains to calculate the small solution when $p_{Lz}>0$. Taking the plus sign in Eq. \eqref{eq:signplzcmsint1}, dividing by the positive $\zeta$ and squaring both (positive) terms we obtain
\begin{equation}
\sign{p_{LzCM}}_{\mathcal{S}>} = \sign{4p^2_{Lz}E_L^2-\left(p^2_{Lz}+E_L^2\right)^2\zeta ^2}\ .
\end{equation}
Plugging in $\zeta^2$, replacing $E_L^2-p_{Lz}^2$ with $M_L^2+p_{\widetilde{T}}^2$ and then factoring $M_L^2+p_{\widetilde{T}}^2$ out gives
\begin{equation}
\sign{p_{LzCM}}_{\mathcal{S}>} = \sign{\frac{\left(E_L^2+p^2_{Lz}\right)^2}{E^2_{LCM}}-\left(M_L^2+p^2_{\widetilde{T}}\right)}\ .
\end{equation}
Since this is the difference of two positive terms, we can take the positive square root of each term.  We can then multiply by the positive $E_{LCM}$, substitute $p_{Lz}^2=E_L^2-M_L^2-p_{\widetilde{T}}^2$, simplify and divide by 2 to get
\begin{equation}
\sign{p_{LzCM}}_{\mathcal{S}>}=\sign{E_L^2-E^2_{L0}(p^2_{\widetilde{T}})} \label{eq:signplzcmsint2},
\end{equation}
with the function $E^2_{L0}(p^2_{\widetilde{T}})$ defined as
\begin{equation}
E^2_{L0}(p^2_{\widetilde{T}}) = \frac{\sqrt{M_L^2+p^2_{\widetilde{T}}}}{2}\left( \sqrt{M_L^2+p^2_{\widetilde{T}}}+E_{LCM}\right).
\end{equation}
We note from Eq.~\eqref{eq:ptmaxancillary} that the maximum value of $p_{\widetilde{T}}$ is given by
\begin{equation}
\sqrt{M_L^2+p^2_{\widetilde{T}max}}=E_{LCM},
\end{equation}
from which it follows that
\begin{equation}
E^2_{L0max}=E^2_{LCM}.
\end{equation}
However, in Appendix~\ref{sec:EL>ELCM} we show that for the small solution when $p_{Lz}>0$, we always have $E_L>E_{LCM}$ giving us
\begin{equation}
\sign{p_{LzCM}}_{\mathcal{S}>}=+1\label{eq:sign pLzCM small plus}\ .
\end{equation}
Putting Eq.~\eqref{eq:sign pLzCM small minus} and \eqref{eq:sign pLzCM small plus} together gives
\begin{equation}
\sign{p_{LzCM}}_{\mathcal{S}}=\sign{p_{Lz}}\ .
\end{equation}
However, we also show in Appendix~\ref{sec:EL>ELCM} that $\sign{p_{Lz}}=\sign{E_L-E_{LCM}}$.  Plugging in Eq.~\eqref{eq:elcms} finally gives us
\begin{equation}
\sign{p_{LzCM}}_{\mathcal{S}} = \sign{E_{L}-\frac{M_B^2+M_L^2-M_D^2}{2M_B}}\ ,
\end{equation}
which completes the derivation of Eq.~\eqref{eq:smallsolution}.  We note that our final solutions depend on only the measured lab-frame energy of $L$ and the known masses. 

\subsection{\boldmath{$p_{\widetilde{T}}$} as a Function of \boldmath{$E_D$}}

We would next like to show that $p_{\widetilde{T}}$ can be calculated directly from the masses of the particles and the energy of $D$.  For this section, we note that Eqs.~\eqref{eq:cosmagfinal} through \eqref{eq:smallsolution} do not depend on our choice of $z$-axis.  Therefore, for this section, we choose a new reference frame for convenience and without loss of generality.  We take the $z$-axis to be along the direction of $\vec{p}_L$.  We further choose the plane of our interaction to be the $x$-$z$ plane.  With this choice, we have
\begin{subequations}
\label{eq:allmomenta}
\begin{eqnarray}
\vec{p}_{L}=\left(0,0,p_{Lz}\right)\ , \\
\vec{p}_{D}=\left( p_{Dx},0,p_{Dz} \right)\ , \\
\vec{p}_{B}=\left( p_{Dx},0,p_{Lz}+p_{Dz} \right)\ .
\end{eqnarray}
\end{subequations}
We can then use the unit vector that is normal to $\vec{p}_B$ in the $x$-$z$ plane,
\begin{equation}
\hat{n}_{\widetilde{T}} = \frac{\left( p_{Lz}+p_{Dz},0,-p_{Dx} \right)}{\sqrt{\left( p_{Lz}+p_{Dz}\right)^2+p_{Dx}^2}}\ ,
\end{equation}
to determine $p_{\widetilde{T}}=\mbox{abs}\left(\vec{p}_L \cdot \hat{n}_{\widetilde{T}}\right)$ giving
\begin{equation}
p^2_{\widetilde{T}} = \frac{p^2_{Dx}p^2_{Lz}}{\left(p_{Lz}+p_{Dz}\right)^2+p^2_{Dx}} \label{eq:pttilde}\ .
\end{equation}
We can use the relation $p_D^2=M_D^2$ to obtain
\begin{equation}
p^2_{Dx}=E_D^2-M_D^2-p^2_{Dz} \label{eq:pdxsquared}\ ,
\end{equation}
at which point, $p_{Dz}$ is the only unknown other than $E_D$.  To obtain this, we expand $M_B^2=\left(p_L+p_D\right)^2$ to get
\begin{equation}
2p_{Lz}p_{Dz}=M_L^2+M_D^2-M_B^2+2E_LE_D \label{eq:plzpdz}\ .
\end{equation}
Using Eqs.~\eqref{eq:pdxsquared} and \eqref{eq:plzpdz}, we can show that the denominator of Eq. \eqref{eq:pttilde} is given by
\begin{equation}
\left(p_{Lz}+p_{Dz}\right)^2+p^2_{Dx} = \left( E_L+E_D \right)^2-M_B^2\ .
\end{equation}
For the numerator of Eq.~\eqref{eq:pttilde}, we can use Eq. \eqref{eq:pdxsquared} to obtain
\begin{equation}
p^2_{Dx}p^2_{Lz} = \left( E_L^2-M_L^2 \right)\left(E_D^2-M_D^2\right)-p^2_{Lz}p^2_{Dz}\ .
\end{equation}
Putting this together, we finally obtain
\begin{equation}
p^2_{\widetilde{T}}=\frac{\left(E_L^2-M_L^2\right)\left(E_D^2-M_D^2\right)-p^2_{Lz}p^2_{Dz}}{\left(E_L+E_D\right)^2-M_B^2}\ ,
\end{equation}
where $p^2_{Lz}p^2_{Dz}$ is given by Eq.~\eqref{eq:plzpdz}. This completes the derivation of Eqs.~\eqref{eq:pTED1} and \eqref{eq:pTED2}.


\section{\label{sec:dylhc}Charged Drell-Yan Lepton Production}

A special case of this kinematic variable was first used in \cite{Chiang:2011kq} where a new resonance in charged Drell-Yan (D-Y) production of a charged lepton and a neutrino was considered.  It was there shown that once the mass of the new resonance is determined, although there is a two-fold ambiguity in the momentum of the neutrino which results in a two-fold ambiguity of the CM frame, it is nevertheless possible to reconstruct the full CM angular distribution of this new resonance.   In this case, the spin and, in some cases, the parity violation of the couplings can be measured from this distribution.  In this section, we will summarize the salient points and discuss the effects of a finite width on these distributions.

\begin{figure}
\includegraphics[scale=.5]{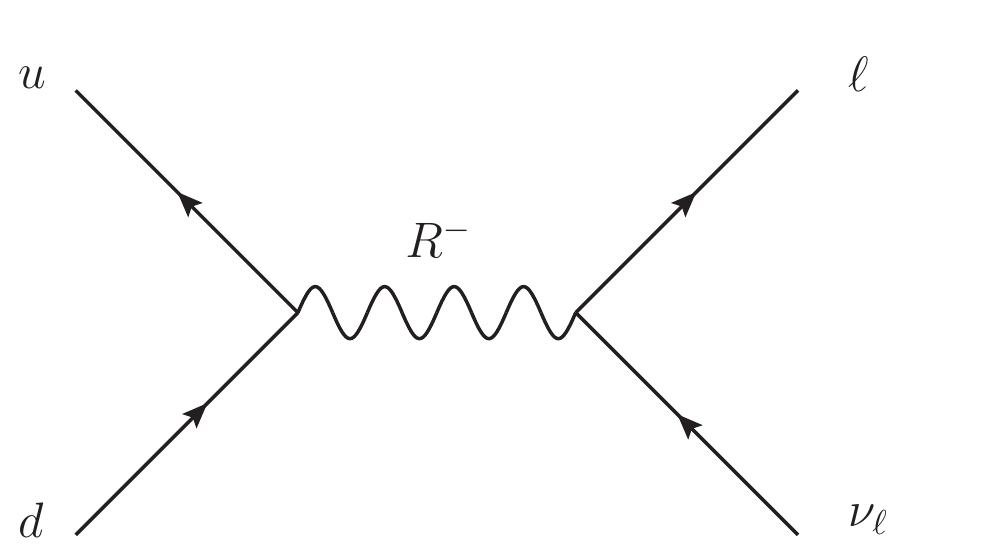}
\caption{\label{fig:dy}The Drell-Yan process $\overline{u}d\rightarrow {\ell}^-{\nu}_{\ell}$, where$R^{-}$ is some general charged resonance.}
\end{figure}

In Figure~\ref{fig:dy}, we show an illustrative diagram for the D-Y resonance producing a charged lepton (an electron or a muon) and a neutrino.  We will call the resonance $R$ and assume its mass has been measured, for example, in a transverse mass distribution.  The mass of both the charged lepton and the neutrino are very small and will be taken as zero for clarity.  Furthermore, due to momentum conservation, the total transverse momentum is zero with the result that the momentum of the charged lepton transverse to $R$ ($p_{\widetilde{T}}$) is equal to its momentum transverse to the beam direction ($p_T$), which is a measured quantity.  As a result, we have
\begin{equation}\label{eq:DY-large solution}
\cos{{\theta}_{\ell R\mathcal{L}}} =-\sqrt{1-\frac{4p_{T}^2}{M_R^2}}
\end{equation}
and
\begin{equation}
\cos{{\theta}_{\ell R\mathcal{S}}}=\sign{E_{\ell}-\frac{M_R}{2}} \sqrt{1-\frac{4p_{T}^2}{M_R^2}}\ .
\end{equation}

In \cite{Chiang:2011kq}, it was shown that the large solution ($\mathcal{L}$) is sufficient to determine the spin of the resonance $R$.  The parity-symmetric angular distribution can be reconstructed by dividing each $\cos{\theta_{\ell R\mathcal{L}}}<0$ bin by $2$ and taking the mirror image for the $\cos{\theta_{\ell R}}>0$ bins.  It was found that in the absence of cuts and for a very small width, the reconstructed angular distribution matched the true distribution perfectly.  It was also shown that the effect of cuts was to remove the large $|\cos{\theta_{\ell R}}|$ bins but did not affect the center of the distribution.  For example, it was shown that a cut of $p_{T}>250$GeV only affected the $|\cos{\theta_{\ell R}}|\gtrsim0.9$ bins.  The effect of a finite width was not explored in \cite{Chiang:2011kq}.  

\begin{figure}
\begin{center}
\includegraphics[scale=1]{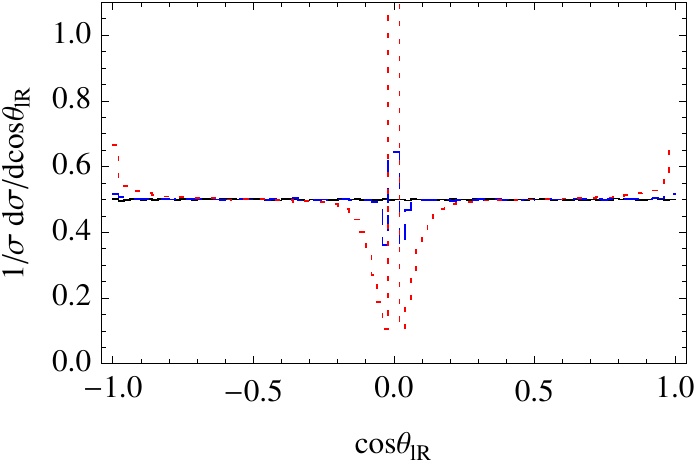}\\
\includegraphics[scale=1]{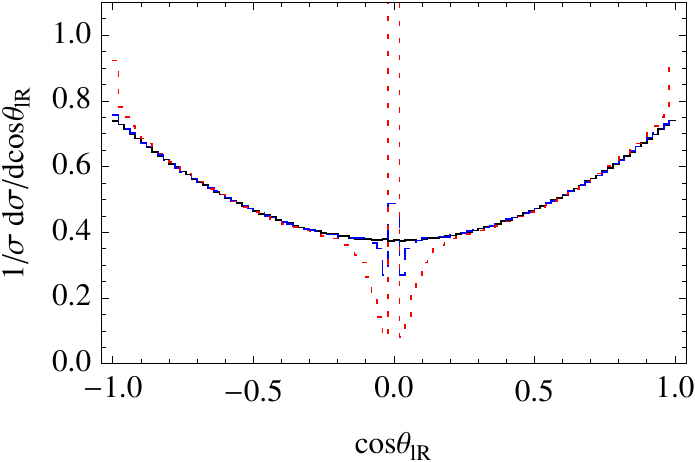}\\
\includegraphics[scale=1]{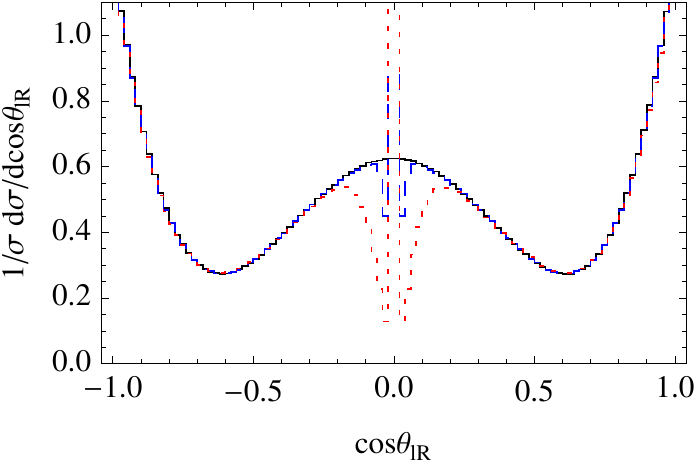}
\end{center}
\caption{\label{fig:DYlarge}CM angular distributions using the large solution (Eq.~\eqref{eq:DY-large solution}) and reconstructing as outlined in the text.  The three curves are solid black for the true CM angular distribution, dashed blue for the reconstructed large solution when the width is 0.1\% of the mass and dotted red when the width is 1\% of the mass.  From top to bottom the three plots are for a scalar, vector and tensor $R$.}
\end{figure}

In Figure~\ref{fig:DYlarge}, we plot the reconstructed CM angular distribution using Eq.~\eqref{eq:DY-large solution} with the effects of a finite width included.  
The three plots are for a scalar (top), vector (middle) and tensor (bottom) resonance, $R$.  For each plot, we have included the true CM angular distribution in solid black, the reconstructed CM angular distribution for a 0.1\% width in dashed blue, and the reconstructed CM angular distribution for a 1\% width in dotted red.  We can see that for widths below 0.1\% of the mass, there is very little alteration of the distribution.  At 1\% of the width, the modification of the distribution is still satisfactory.  Only the bins for $|\cos{\theta_{\ell R}}|<0.2$ and $|\cos{\theta_{\ell R}}|>0.9$ are significantly affected.  However, for widths above 1\% of the mass, the modification of the distribution starts becoming significant.

The small solution ($\mathcal{S}$) can be used to reconstruct the CM angular distribution as well, following the procedure outlined in \cite{Chiang:2011kq}.  Although the reconstruction method is slightly more complicated, it has the advantage that the parity violation can also be reconstructed.  The effect of acceptance cuts was discussed in \cite{Chiang:2011kq} and is similar to that of the large solution.  We have analyzed the effect of finite widths and find it to be similar to the large solution case presented here.

\section{\label{sec:antlers}Antlers at the ILC}
\begin{figure}
\begin{center}
\includegraphics[scale=0.35]{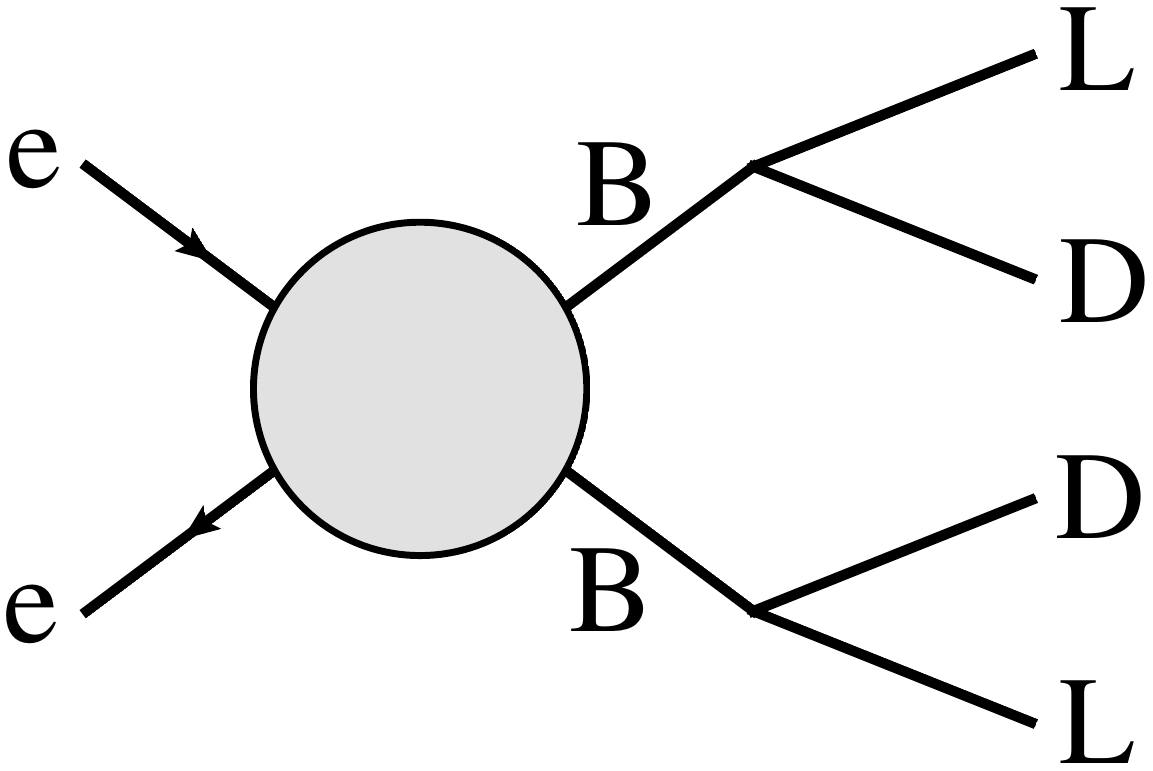}
\end{center}
\caption{\label{fig:antler}An antler diagram for the pair production of two $B$ fields which then each decay to $L$ and $D$.  The energy of each $D$ is known in this case as described in the text.}
\end{figure}

We will now consider diagrams of the type shown in Fig.~\ref{fig:antler}.  We will call this type of diagram an ``antler" diagram, following \cite{Han:2009ss,Christensen:2014yya} where the authors showed that the mass of $B$ and $D$ could, in principle, be measured in this process.  
An analysis of this diagram shows that, given the masses of $B$ and $D$, there is a two-fold ambiguity in the momenta of the two $D$s (see App.~\ref{sec:twofold ambiguity}).  Nevertheless, the energy of the two $D$s are known.  This can be seen by noting that
\begin{equation}
E_{B_i} = E_B = \frac{\sqrt{s}}{2}\ ,
\end{equation}
where $\sqrt{s}$ is the collision energy and $i$ determines which $B$ is being referred to.  Conservation of energy, then, gives us
\begin{equation}
E_{D_i} = E_B - E_{L_i} = \frac{\sqrt{s}}{2} - E_{L_i}\ .
\end{equation}
Since $E_{L_i}$ is a measured quantity, $E_{D_i}$ is known (where $i$ determines which $L$ and $D$ is being considered).  As we saw in Sec.~\ref{sec:derivation}, this is sufficient to apply our methods to this process, even though the $B_i$ CM frame can not be reconstructed.

\subsection{\label{sec:Antlers-muons}\boldmath{$e^+e^-\to\gamma^*/Z^*\to B^+B^-\to\mu^+\mu^-DD$}}
\begin{figure}
\includegraphics[scale=.5]{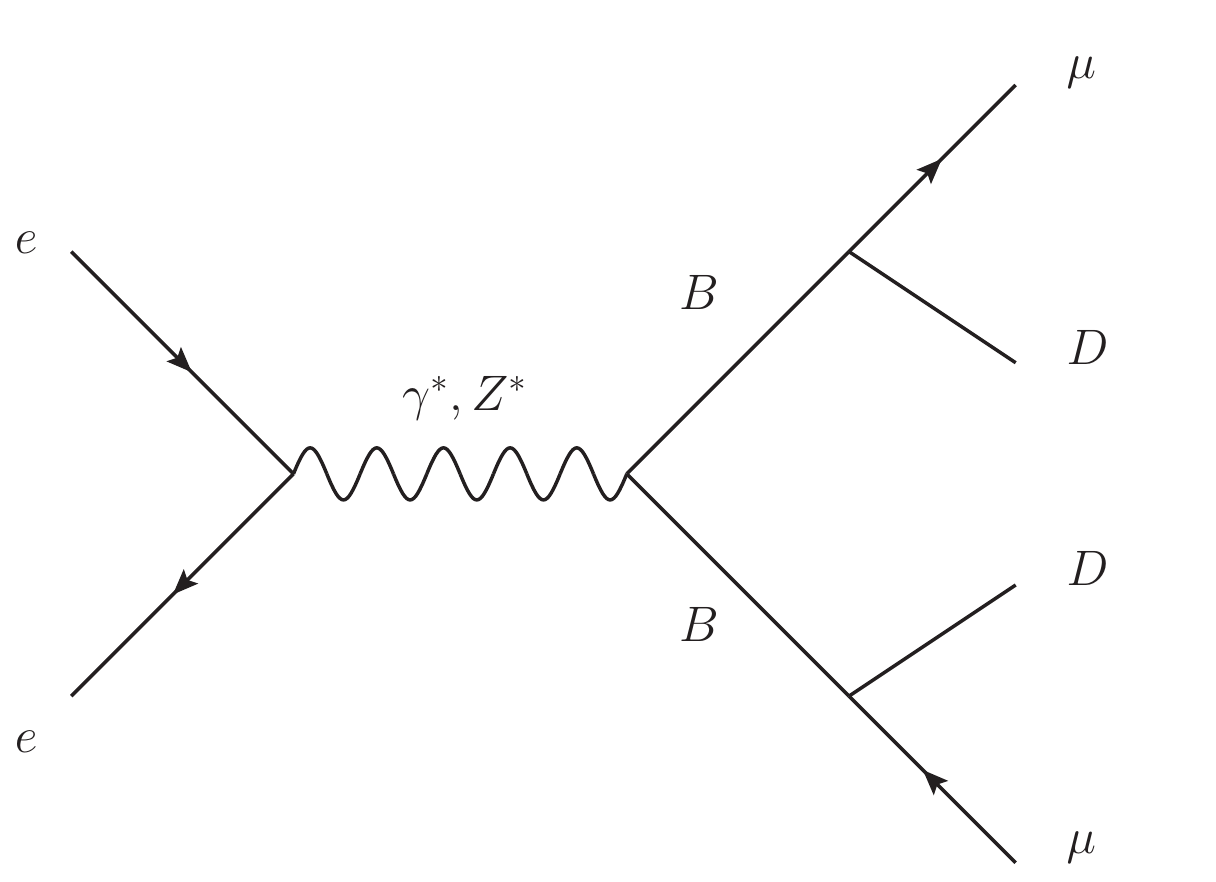}
\caption{\label{fig:ilcschannelmuon}Pair production of $B$ followed by its decay to a muon and  $D$ at the ILC.}\end{figure}

For definiteness, we will take $L$ to be a muon and allow $B$ and $D$ to take any consistent spin between $0$ and $2$.  The spins and charges of $B$ and $D$ are given in Table~\ref{table:particlenames}.  
\begin{table}
\begin{tabular}{|c|ccccc|c|}
\hline
 Particle & & & Spin & & & Charge \\
 & 0 & 1/2 & 1 & 3/2 & 2 & \\
\hline\hline
$\ell$ & & $\ell$ & & & & $-1$ \\
$D$ & $\sD$ & $\fD$ & $\vD$ & $\rD$ & $\tD$ & 0  \\
$B$ & $\sB$ & $\fB$ & $\vB$ & $\rB$ & $\tB$ & $-1$\\
\hline
\end{tabular}
\caption{\label{table:particlenames}List of symbols used for the particles in our analysis with their charges and spins.}
\end{table}
We will further assume that $D$ is self-charge-conjugate.
We will take the two $B$s to be produced by their interaction with the photon and $Z$ boson as in Fig.~\ref{fig:ilcschannelmuon}.
For this first illustrative use of our new kinematical variable at the ILC, we will take all new interactions to be parity symmetric.  The parity violating case will be considered in a later publication.
We used FeynRules \cite{Christensen:2008py,Christensen:2009jx,Christensen:2013aua,Alloul:2013bka} to implement these operators into CalcHEP \cite{Belyaev:2012qa} format.  We implemented our kinematic observables (Eqs.~\eqref{eq:cosmagfinal} through \eqref{eq:pTED2}) using the usrmod functionality of CalcHEP.  We then performed extensive simulations of all the consistent spin-combinations.

We note that the parity-conserving operators split into two categories.  The first category gives a symmetric CM angular distribution which only depends on the masses and the collision energy.  All dependence on the coupling constants drops out of the normalized differential cross-sections.  The majority of the parity-conserving operators fall into this category and were included in our analysis, in a model-independent way.  The resulting on-shell differential cross-sections were calculated analytically.  They have been included in Table~\ref{tab:analytic CM angular distributions} and will be discussed further below.  A complete list of the operators in this category can be found in App.~\ref{sec:effective operators}.  A discussion of the derivation of the analytic differential cross-sections coming from these operators is included in Appendix~\ref{sec:CM analytic derivation}.  

The second category of parity-conserving operators also gives a symmetric CM angular distribution, however the expressions for the differential cross-sections are more complicated and depend on the values of the coupling constants.  In particular, the values of $\mathcal{A}$, $\mathcal{B}$, $\mathcal{C}$ and $\mathcal{N}$ in Table~\ref{tab:analytic CM angular distributions} depend non-trivially on the coupling constants in the second category of parity-conserving operators (a discussion of the operators from this category is also included in Appendix~\ref{sec:effective operators}).   Since our main purpose in the present article is to show that our kinematical variables faithfully reproduce the true CM angular distribution, for simplicity, we focus on the parity-conserving operators from the first category.

In order to ascertain the power of our kinematic variables, we would like to compare its results with the ``true" CM angular distribution.  Since the two $D$s in this process are identical, we numerically calculated the true CM angular distribution in two ways.  First, in the process $e^+e^-\to B^+B^-\to\mu^+\mu^-DD$, we used the Monte Carlo information to determine which $D$ to pair with $\mu^-$.  We then boosted into this frame to calculate the true CM angular distribution.  Second, we simulated the process $e^+e^-\to\mu^-DB^+$, where the final $B^+$ was not decayed.  We then boosted into the $\mu^-D$ rest frame and determined the CM angular distribution.  In all cases, we found agreement between these methods, as expected.

\begin{figure*}
\begin{center}
\begin{minipage}{1.7in}
\includegraphics[scale=0.8]{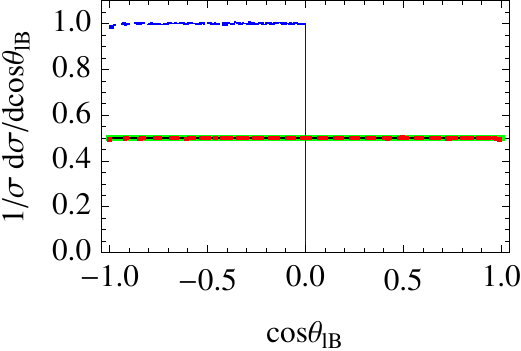}\\
(i)
\end{minipage}
\begin{minipage}{1.7in}
\includegraphics[scale=0.8]{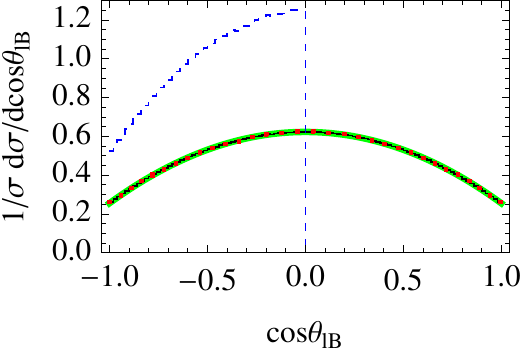}\\
(ii)
\end{minipage}
\begin{minipage}{1.7in}
\includegraphics[scale=0.8]{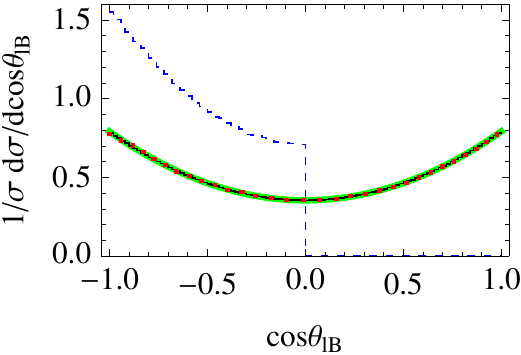}\\
(iiia)
\end{minipage}
\begin{minipage}{1.7in}
\includegraphics[scale=0.8]{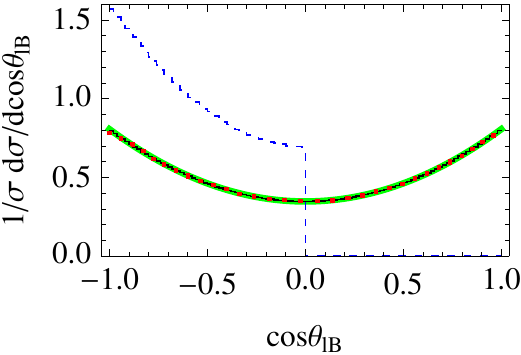}\\
(iiib)
\end{minipage}\\
\vspace{0.1in}
\begin{minipage}{1.7in}
\includegraphics[scale=0.8]{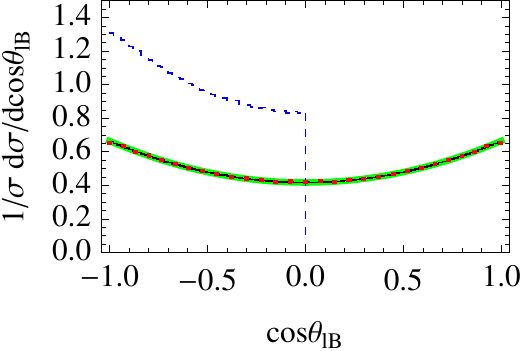}\\
(iiic)
\end{minipage}
\begin{minipage}{1.7in}
\includegraphics[scale=0.8]{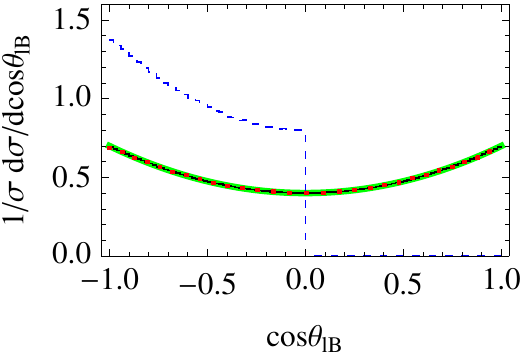}\\
(iiid)
\end{minipage}
\begin{minipage}{1.7in}
\includegraphics[scale=0.8]{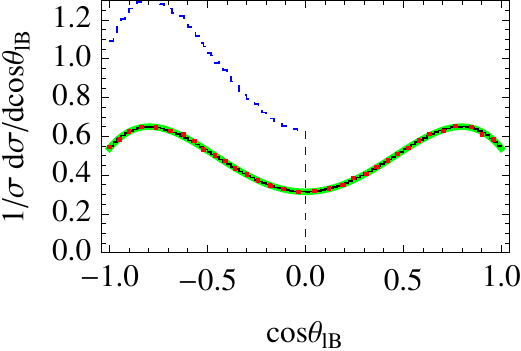}\\
(iva)
\end{minipage}
\begin{minipage}{1.7in}
\includegraphics[scale=0.8]{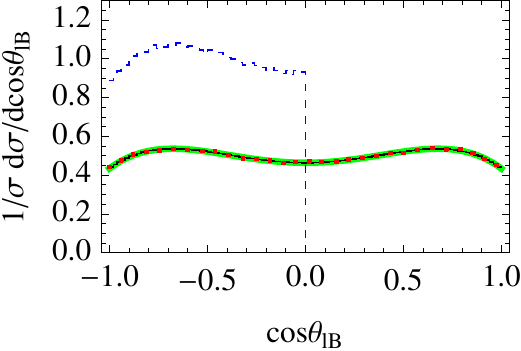}\\
(ivb)
\end{minipage}
\caption{\label{fig:classes plots}The CM angular distributions described in the text.  The solid green lines come from the analytic formulas given in Table~\ref{tab:analytic CM angular distributions}, the solid black lines come from the true simulated CM angular distributions, the dashed blue lines come from the large solution given in Eq.~\ref{eq:largesolution}, and the dotted red lines come from dividing the dashed blue lines in half and taking the mirror image on the $\cos{\theta_{\ell B}}>0$ side.  We used $\sqrt{s}=500$GeV, $M_B=200$GeV and $M_D=50$GeV for this figure.}
\end{center}
\end{figure*}

We simulated the \textit{true} CM angular distribution for every consistent spin-combination.
In all narrow width cases in the absence of cuts, we found perfect agreement between the analytic expressions from Table~\ref{tab:analytic CM angular distributions} and the simulations of the true CM angular distributions. This agreement can be seen in Fig.~\ref{fig:classes plots} where the analytic formulas are used to produce the solid green curves and the Monte Carlo simulation of the \textit{true} solution is plotted in solid black.  The illustrative values $\sqrt{s}=500$GeV, $M_B=200$GeV and $M_D=50$GeV were used to create this figure.  
As expected, all the angular distributions are symmetric since we only considered parity-symmetric operators in this analysis.
\begin{table*}
\begin{tabular}{llll}
 & $s_B$ & $s_D$ & \multicolumn{1}{c}{$\frac{1}{\sigma}\frac{d\sigma}{d\cos{\theta_{\ell B}}}$}\\
\hline
i & $0$ & & \multicolumn{1}{c}{$\frac{1}{2}$}\\
\hline
i & $\frac{1}{2}$ & & \multicolumn{1}{c}{$\frac{1}{2}$}\\
\hline
ii & $1$ & $\frac{1}{2}$ & \multicolumn{1}{c}{$\frac{\mathcal{A}_{ii}}{\mathcal{N}_{ii}}-\frac{\mathcal{B}_{ii}}{\mathcal{N}_{ii}}\cos^2\theta_{\ell B}$}\\
\\
&&& $\mathcal{A}_{ii} = 3M_B^2\left(s^2-4M_B^2s+4M_B^2M_D^2+8M_B^4\right)$\\
&&& $\mathcal{B}_{ii} = 3s\left(M_B^2-M_D^2\right)\left(s-4M_B^2\right)$\\
&&& $\mathcal{N}_{ii} = 2\left(2M_B^2+M_D^2\right)\left(s^2-4M_B^2s+12M_B^4\right)$\\
\hline
iiia & $1$ & $\frac{3}{2}$ & \multicolumn{1}{c}{$\frac{\mathcal{A}_{iiia}}{\mathcal{N}_{iiia}}+\frac{\mathcal{B}_{iiia}}{\mathcal{N}_{iiia}}\cos^2\theta_{\ell B}$}\\
\\
&&& $\mathcal{A}_{iiia} = 12M_B^2\left(M_D^2s^2-4M_B^2M_D^2s+M_B^6+10M_B^4M_D^2+M_B^2M_D^4\right)$\\
&&& $\mathcal{B}_{iiia} = 3s\left(M_B^2-M_D^2\right)^2\left(s-4M_B^2\right)$\\
&&& $\mathcal{N}_{iiia} = 2\left(M_B^4+10M_B^2M_D^2+M_D^4\right)\left(s^2-4M_B^2s+12M_B^4\right)$\\
\hline
iiib & $\frac{3}{2}$ & $0$ & \multicolumn{1}{c}{$\frac{\mathcal{A}_{iiib}}{\mathcal{N}_{iiib}}+\frac{\mathcal{B}_{iiib}}{\mathcal{N}_{iiib}}\cos^2\theta_{\ell B}$}\\
\\
&&& $\mathcal{A}_{iiib} = s^3-2M_B^2s^2+4M_B^4s+72M_B^6$\\
&&& $\mathcal{B}_{iiib} = 3s\left(s-4M_B^2\right)\left(s+2M_B^2\right)$\\
&&& $\mathcal{N}_{iiib} = 4\left(s^3-2M_B^2s^2-2M_B^4s+36M_B^6\right)$\\
\hline
iiic & $\frac{3}{2}$ & $1$ & \multicolumn{1}{c}{$\frac{\mathcal{A}_{iiic}}{\mathcal{N}_{iiic}}+\frac{\mathcal{B}_{iiic}}{\mathcal{N}_{iiic}}\cos^2\theta_{\ell B}$}\\
\\
&&& $\mathcal{A}_{iiic} = \left(s^3-2M_B^2s^2+4M_B^4s+72M_B^6\right)\left(M_B^4+10M_B^2M_D^2+M_D^4\right)+12M_B^2M_D^2s\left(s+2M_B^2\right)\left(s-4M_B^2\right)$\\
&&& $\mathcal{B}_{iiic} = 3s\left(M_B^2-M_D^2\right)^2\left(s-4M_B^2\right)\left(s+2M_B^2\right)$\\
&&& $\mathcal{N} _{iiic}= 4\left(M_B^4+10M_B^2M_D^2+M_D^4\right)\left(s^3-2M_B^2s^2-2M_B^4s+36M_B^6\right)$\\
\hline
iiid & $\frac{3}{2}$ & $2$ & \multicolumn{1}{c}{$\frac{\mathcal{A}_{iiid}}{\mathcal{N}_{iiid}}+\frac{\mathcal{B}_{iiid}}{\mathcal{N}_{iiid}}\cos^2\theta_{\ell B}$}\\
\\
&&& $\mathcal{A}_{iiid} = \left(2M_B^6+47M_B^4M_D^2+128M_B^2M_D^4+3M_D^6\right)\left(s^3-2M_B^2s^2-2M_B^4s+36M_B^6\right)$\\
&&& \hspace{0.5in}$+6M_B^4\left(2M_B^6+11M_B^4M_D^2-16M_B^2M_D^4+3M_D^6\right)\left(s+6M_B^2\right)$\\
&&& $\mathcal{B}_{iiid} = 3s\left(2M_B^6+11M_B^4M_D^2-16M_B^2M_D^4+3M_D^6\right)\left(s-4M_B^2\right)\left(s+2M_B^2\right)$\\
&&& $\mathcal{N}_{iiid} = 4\left(2M_B^6+29M_B^4M_D^2+56M_B^2M_D^4+3M_D^6\right)\left(s^3-2M_B^2s^2-2M_B^4s+36M_B^6\right)$\\
\hline
iva & $2$ & $\frac{1}{2}$ & \multicolumn{1}{c}{$\frac{\mathcal{A}_{iva}}{\mathcal{N}_{iva}} + \frac{\mathcal{B}_{iva}}{\mathcal{N}_{iva}}\cos^2\theta_{\ell B} - \frac{\mathcal{C}_{iva}}{\mathcal{N}_{iva}}\cos^4\theta_{\ell B}$}\\
\\
&&& $\mathcal{A}_{iva} = 5\left(M_D^2s^4-8M_B^2M_D^2s^3+28M_B^4M_D^2s^2+27M_B^{6}s^2-48M_B^6M_D^2s-108M_B^8s+144M_B^8M_D^2+216M_B^{10}\right)$\\
&&& $\mathcal{B}_{iva} = -15s\left(s-4M_B^2\right)\left(-3M_B^2s^2+2M_D^2s^2-8M_B^2M_D^2s+12M_B^4s-12M_B^4M_D^2-9M_B^6\right)$\\
&&& $\mathcal{C}_{iva} = 45s^2\left(M_B^2-M_D^2\right)\left(s-4M_B^2\right)^2$\\
&&& $\mathcal{N}_{iva} = 4\left(3M_B^2+2M_D^2\right)\left(s^4-8M_B^2s^3+46M_B^4s^2-120M_B^6s+180M_B^8\right)$\\
\hline
ivb & $2$ & $\frac{3}{2}$ & \multicolumn{1}{c}{$\frac{\mathcal{A}_{ivb}}{\mathcal{N}_{ivb}} + \frac{\mathcal{B}_{ivb}}{\mathcal{N}_{ivb}}\cos^2\theta_{\ell B} - \frac{\mathcal{C}_{ivb}}{\mathcal{N}_{ivb}}\cos^4\theta_{\ell B}$}\\
\\
&&& $\mathcal{A}_{ivb} = 889s^4-7112M_B^2s^3+49804M_B^4s^2-142320M_B^6s+216720M_B^8$\\
&&& $\mathcal{B}_{ivb} = 270s\left(s-4M_B^2\right)\left(11s^2-44M_B^2s+6M_B^4\right)$\\
&&& $\mathcal{C}_{ivb} = 3375 s^2\left(s-4M_B^2\right)^2$\\
&&& $\mathcal{N}_{ivb} = 2408\left(s^4-8M_B^2s^3+46M_B^4s^2-120M_B^6s+180M_B^8\right)$\\
\hline
\end{tabular}
\caption{\label{tab:analytic CM angular distributions}The analytic expressions for the CM angular distributions for each spin combination.  The first column gives the class described in the text and shown in Fig.~\ref{fig:classes plots}, the column labeled $s_B$ gives the spin of $B$, the column labeled $s_D$ gives the spin of $D$ and the column labeled $1/\sigma\ d\sigma/d\cos\theta_{\ell B}$ gives the CM angular distribution.  See Appendix~\ref{sec:CM analytic derivation} for a discussion of the derivation of these formulas which included the photon, Z-boson and interference diagrams.}
\end{table*}

We also simulated the full signal final state $\mu^+\mu^-DD$ and calculated our kinematical variable using Eqs.~\eqref{eq:cosmagfinal} through \eqref{eq:pTED2} using \textit{only} the momentum of the muon.  The large solution (Eq.~\ref{eq:largesolution}) is shown as the dashed blue line in Fig.~\ref{fig:classes plots}.  As expected, it is nonzero only for $\cos\theta_{\ell B}<0$ and exactly double the true distribution.  We can reconstruct the true solution by dividing this distribution by $2$ and taking the mirror image on the $\cos\theta_{\ell B}>0$ side.  This has been done and is plotted as the red dotted curves in Fig.~\ref{fig:classes plots}.  As can be seen, the red dotted curves agree perfectly with the true solutions in solid green and solid black.  The small solution would give a similar result, but would require a slightly more complicated reconstruction method.  However, because the distribution is fundamentally symmetric in the present article, the sign is not helpful.  Therefore, since the large solution is sufficient to get exact agreement with the true solution, we will not discuss the small solution further in this section.

\begin{figure}
\begin{center}
\includegraphics[scale=1]{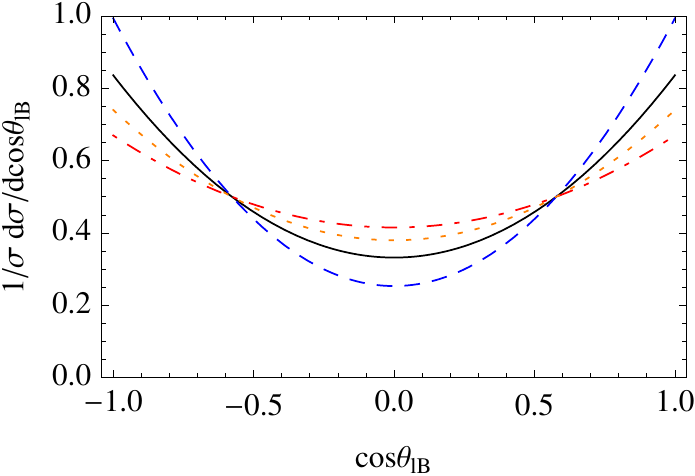}\\
(iii)\\
\includegraphics[scale=1]{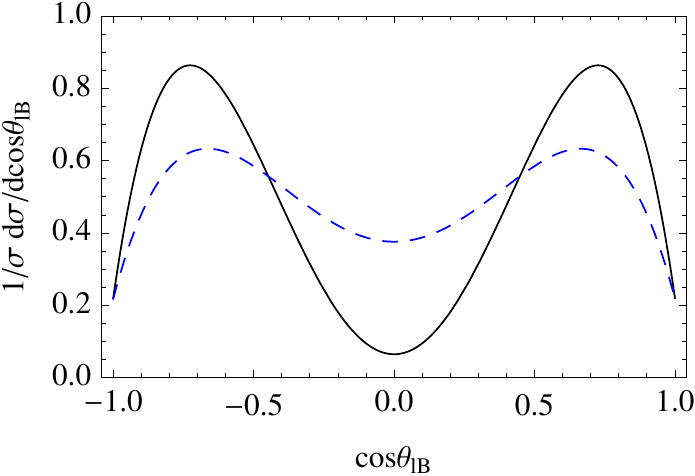}\\
(iv)
\end{center}
\caption{\label{fig:iii splitting}Class iii and iv CM angular distributions for the values $\sqrt{s}=1$TeV, $M_B=200$GeV and $M_D=70$GeV.  In the case of iii, the solid black line is for class iiia, the blue dashed line is for class iiib, the red dot-dashed line is for class iiic and the orange dotted line is for class iiid.  In the case of iv, the solid black line is for case iva while the dashed blue line is for case ivb.}
\end{figure}

We have found that the various spin-combinations fall into four major classes: i) flat distributions, ii) parabolic distributions with negative concavity, iii) parabolic distributions with positive concavity, and iv) ``M" shaped distributions.  This classification corresponds with the first column of Table~\ref{tab:analytic CM angular distributions} and the labels in Figs.~\ref{fig:classes plots} through \ref{fig:ISR plots} and described in detail below.

The first class is given in row (i) of Table~\ref{tab:analytic CM angular distributions} and shown in Fig.~\ref{fig:classes plots}i.  It is a flat line with no $\cos\theta_{\ell B}$ dependence.  This class not only includes the case where $B$ has spin $0$, but also includes the cases where $B$ has spin $\frac{1}{2}$.  The reason for this is that the inherent parity symmetry of these angular distribution requires equal contributions from $d^j_{m,+m'}$ and $d^j_{m,-m'}$ for any given $j$, $m$ and $m'$.  In the case of a spin-$\frac{1}{2}$ $B$, the Wigner $d^j_{mm'}$-functions are $d^{\frac{1}{2}}_{\frac{1}{2},\frac{1}{2}}\left(\theta_{\ell B}\right)=\cos\left(\theta_{\ell B}/2\right)$ and $d^{\frac{1}{2}}_{\frac{1}{2},-\frac{1}{2}}\left(\theta_{\ell B}\right)=-\sin{\left(\theta_{\ell B}/2\right)}$.  As a result, since the differential cross section is proportional to the linear combination of the squares of the Wigner $d^j_{mm'}$-functions (see App.~\ref{sec:Wigner d-functions}), the normalized differential cross section in this case is given by
\begin{equation}
\frac{1}{\sigma_{i}}\frac{d\sigma_{i}}{d\cos{\theta_{\ell B}}} =
\frac{1}{2}\left[\left(d^{\frac{1}{2}}_{\frac{1}{2},\frac{1}{2}}\left(\theta_{\ell B}\right)\right)^2+\left(d^{\frac{1}{2}}_{\frac{1}{2},-\frac{1}{2}}\left(\theta_{\ell B}\right)\right)^2\right] =
 \frac{1}{2}\ .
 \end{equation}

The second class of CM angular distributions is given by a concave negative parabola as seen in Fig.~\ref{fig:classes plots}ii.  There is only one spin combination that gives this distribution and that is a spin-$1$ $B$ decaying to a spin-$\frac{1}{2}$ $D$.  The Wigner d-functions for spin-$1$ are $d^1_{1,\pm1}\left(\theta_{\ell B}\right)=\frac{1}{2}\left(1\pm\cos\theta_{\ell B}\right)$, $d^1_{1,0}\left(\theta_{\ell B}\right)=-\frac{1}{\sqrt{2}}\sin\theta_{\ell B}$ and $d^1_{0,0}\left(\theta_{\ell B}\right)=\cos\theta_{\ell B}$, whose squares are quadratic in $\cos\theta_{\ell B}$, resulting in a parabola.  The expression for the normalized differential cross section in this case is given in row (ii) of Table~\ref{tab:analytic CM angular distributions}.  We see that the concavity is proportional to $\left(M_B^2-M_D^2\right)\left(s-4M_B^2\right)$ and that, while always negative in this case, is increased in size as the mass difference between $B$ and $D$ is increased and also as the difference between the collision energy $\sqrt{s}$ and the mass of $B$ is increased.  Although we do not have any control over the mass difference between $B$ and $D$, we do have some control over the collision energy.  If a distribution appears to be nearly flat, the collision energy should be increased to determine whether this property changes.  This dependence of the shape on the collision energy is a feature of all the classes discussed here except, of course, class i, the flat distribution.

The third class of distributions is given by concave up parabolas as shown in Fig.~\ref{fig:classes plots}iii.  There are four spin-combinations that give this class of distributions.  The first is a spin-$1$ $B$ decaying to a spin-$\frac{3}{2}$ $D$ (see Fig.~\ref{fig:classes plots}iiia).  The second, third and fourth are, respectively, a spin-$\frac{3}{2}$ $B$ decaying to a spin-$0$ $D$ (see Fig.~\ref{fig:classes plots}iiib), a spin-$1$ $D$ (see Fig.~\ref{fig:classes plots}iiic) and a spin-$2$ $D$ (see Fig.~\ref{fig:classes plots}iiid).  Initially, when $B$ is spin-$\frac{3}{2}$, the Wigner d-functions squared are cubic in $\cos\theta_{\ell B}$. However, since these distributions are inherently parity symmetric, the odd terms cancel (as in the previously discussed spin-$\frac{1}{2}$ case) and we are left with expressions quadratic in $\cos\theta_{\ell B}$.  We further note that, although the Fig.~\ref{fig:classes plots}iiia distribution appears to be equal to the Fig.~\ref{fig:classes plots}iiib distribution and the Fig.~\ref{fig:classes plots}iiic distribution appears to be equal to the Fig.~\ref{fig:classes plots}iiid distribution, this is only a coincidence of the illustrative masses and collision energies we used.  If we, instead, used $\sqrt{s}=1$TeV, $M_B=200$GeV and $M_D=70$GeV, the four distributions would all be separated as in Fig.~\ref{fig:iii splitting}iii.
As we see in Table~\ref{tab:analytic CM angular distributions}, in each of these cases, the strength of the concavity is dependent on both the mass splitting between $B$ and $D$ as well as on the collision energy.  We cannot choose the mass splitting.  However, whatever the mass splitting, we can, in principle, choose higher collision energies to accentuate the shape of each of these cases.

The final class of distributions is given by an ``M" shape which is quartic in $\cos\theta_{\ell B}$.  This comes from a spin-$2$ $B$ decaying to either a spin-$\frac{1}{2}$ $D$ (see Fig.~\ref{fig:classes plots}iva) or a spin-$\frac{3}{2}$ $D$ (see Fig.~\ref{fig:classes plots}ivb).  As in the previous classes, the quadratic and quartic terms in the angular dependence (see rows iva and ivb in Table~\ref{tab:analytic CM angular distributions}) are proportional to $\left(s-4M_B^2\right)$, therefore, the shape can be accentuated by increasing the collision energy.  We give an example of this in Fig.~\ref{fig:iii splitting}iv, where $\sqrt{s}=1$TeV, $M_B=200$GeV and $M_D=70$GeV.


\subsection{\label{sec:finite width}Finite Width Effects}
\begin{figure*}
\begin{center}
\begin{minipage}{1.7in}
\includegraphics[scale=0.8]{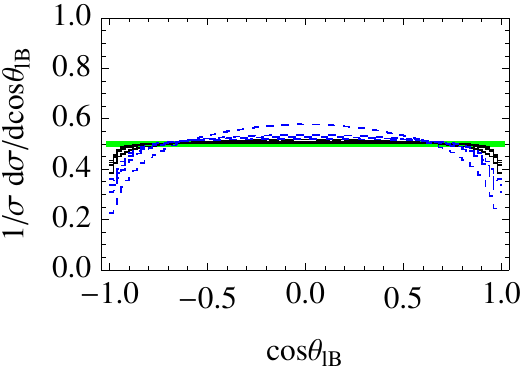}\\
(i)
\end{minipage}
\begin{minipage}{1.7in}
\includegraphics[scale=0.8]{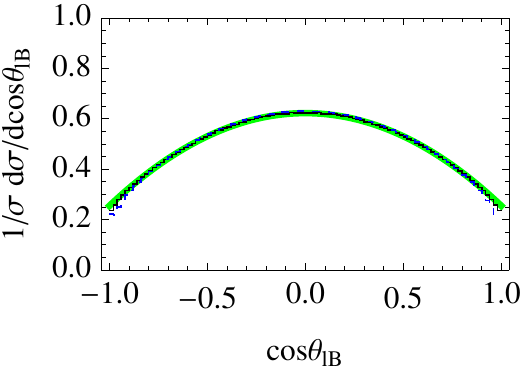}\\
(ii)
\end{minipage}
\begin{minipage}{1.7in}
\includegraphics[scale=0.8]{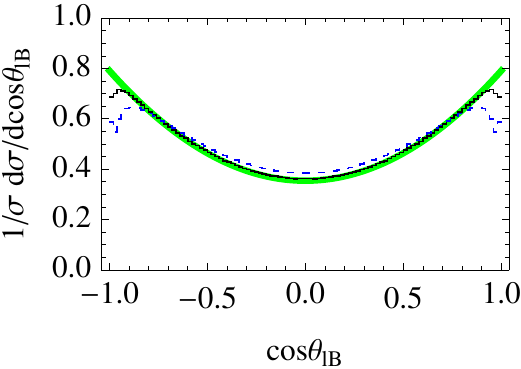}\\
(iiia)
\end{minipage}
\begin{minipage}{1.7in}
\includegraphics[scale=0.8]{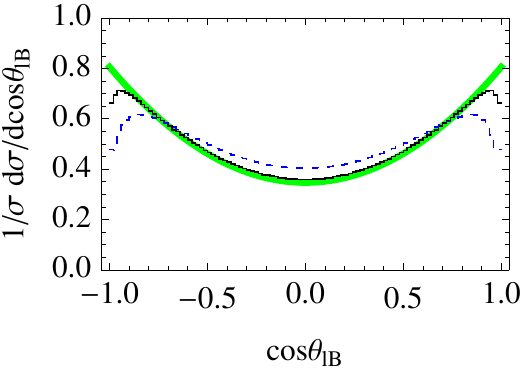}\\
(iiib)
\end{minipage}\\
\vspace{0.1in}
\begin{minipage}{1.7in}
\includegraphics[scale=0.8]{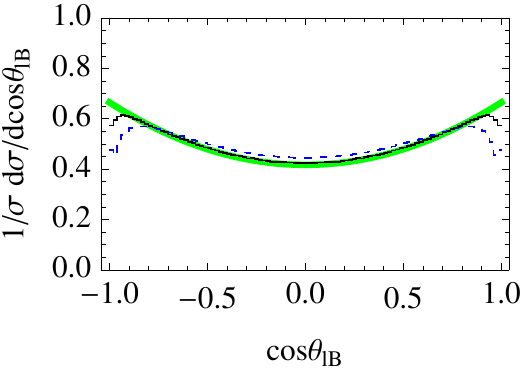}\\
(iiic)
\end{minipage}
\begin{minipage}{1.7in}
\includegraphics[scale=0.8]{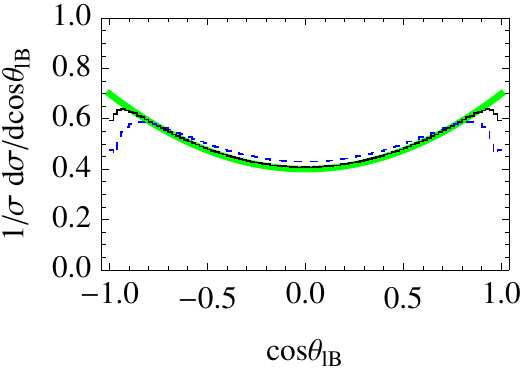}\\
(iiid)
\end{minipage}
\begin{minipage}{1.7in}
\includegraphics[scale=0.8]{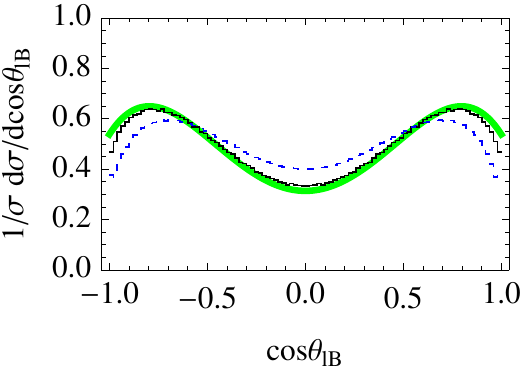}\\
(iva)
\end{minipage}
\begin{minipage}{1.7in}
\includegraphics[scale=0.8]{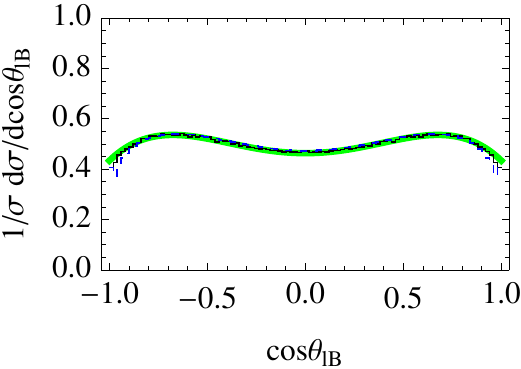}\\
(ivb)
\end{minipage}
\caption{\label{fig:FW plots}The effect of a finite width on the CM angular distributions.  The solid green lines come from the on-shell angular distributions given by the analytic formulas in Table~\ref{tab:analytic CM angular distributions}, the solid black and dashed blue lines come from the large solution given in Eq.~\eqref{eq:largesolution} where the width of $B$ is taken to be $1$\% and $5$\%, respectively.  The collision energy and masses are the same as in Fig.~\ref{fig:classes plots}.}
\end{center}
\end{figure*}

In the previous subsection, we took the width of $B$ to be very small ($0.05$\% of the mass) and found perfect agreement between the \textit{true} CM angular distributions and our reconstructed distributions.  However, the derivation of our kinematical variables, Eqs.~\eqref{eq:cosmagfinal} through \eqref{eq:pTED2}, relied heavily on $B$ being on-shell, and therefore having a narrow width.  Furthermore, off-shell, even the true CM angular distribution could be modified by off-shell polarization vectors.  When the spin of B is greater than 1, the off-shell polarization effect is model dependent.  In this subsection, we would like to estimate the total effect on our reconstruction technique of having a more sizable width.  In Fig.~\ref{fig:FW plots}, we have plotted our reconstructed CM angular distribution when the width of $B$ is $1$\% in solid black and $5$\% in dashed blue.  For comparison, we have also plotted the \textit{true} on-shell CM angular distribution in solid green.  For spins greater than 1, we have used the standard propagators that are built into CalcHEP \cite{Belyaev:2012qa,Christensen:2013aua}.

In general, we find that the effect of the width is almost negligible when $1$\% of the mass and only begins to be sizable for some spin combinations when approximately $5$\% of the mass.  However, even at $5$\%, the agreement with the theoretical curve is still quite good in most cases.  The main effect is to reduce the CM angular distribution for large $|\cos\theta_{\ell B}|$ and, to a lesser extent, to increase the distribution for small $|\cos\theta_{\ell B}|$.  In the case of the theoretically concave negative distribution (see Fig.~\ref{fig:FW plots}ii), the effect of both a $1$\% and $5$\% width is negligible.

In the case of the theoretically flat distribution (see Fig.~\ref{fig:FW plots}i), the effect of a $1$\% width is to slightly reduce the edges when $|\cos\theta_{\ell B}|\gtrsim0.95$ while the enhancement in the center is not very great.  When the width becomes $5$\%, the effect begins to be significant but depends on which spin combination is being considered.  The distributions for spin-$0$ $B$ and for spin-$\frac{1}{2}$ $B$ decaying to a spin-$0$ or spin-$1$ $D$, the effect is still small with the edges only being strongly affected for $|\cos\theta_{\ell B}|\gtrsim0.85$ and the center not being strongly enhanced.  However, for a spin-$\frac{1}{2}$ $B$ decaying to a spin-$2$ $D$, the effect is much larger with the shape changing significantly to look nearly like a parabola with negative concavity.  However, since the negative concavity parabola (spin-$1$ $B$ decaying to a spin-$\frac{1}{2}$ $D$) is not significantly affected by a finite width, it should not be difficult to distinguish these cases.

In the case of the concave positive distributions (see Fig.~\ref{fig:FW plots}iii), the effect is small and mainly at the edges.  For a $1$\% width, the reduction is visible for $|\cos\theta_{\ell B}|\gtrsim0.95$ while the central enhancement is minimal.  For a $5$\% width, the reduction is visible for $|\cos\theta_{\ell B}|\gtrsim0.85$.  Again, the enhancement is minimal.

For the case of a spin-$2$ $B$ decaying to a spin-$\frac{1}{2}$ $D$ (see Fig.~\ref{fig:FW plots}iva), the effect of a $1$\% width is minimal, however, a $5$\% width does change the distribution.  Although it retains the ``M" shape, the outer edges are approximately $20$\% lower while the center is approximately $20$\% higher.  In the case of a spin-$2$ $B$ decaying to a spin-$\frac{3}{2}$ $D$ (see Fig.~\ref{fig:FW plots}ivb), the effect of even a $5$\% width is very small with a slight decrease towards the edges.


\subsection{\label{sec:ISR}ISR, Beamstrahlung and Cuts}
We also considered the effects of initial state radiation (ISR), beamstrahlung and basic cuts in an electron-positron collider environment on our distributions.  The cuts we used were 
\begin{eqnarray}\label{eq:cos cut}
&|\cos\theta_l|<0.9962&\\
&E_l>10\mbox{GeV}&
\end{eqnarray}
to ensure detection, where $\theta_l$ and $E_l$ are, respectively, the polar angle and the energy of the electron in the lab frame, and 
\begin{equation}\label{eq:pT cut}
\slashed{p}_{T}>10\mbox{GeV}
\end{equation}
to remove the photon T-channel induced background from $e^+e^-\to e^+e^-\mu^+\mu^-$ where the final electron-positron pair are missed \cite{Christensen:2014yya}, where $\slashed{p}_T$ is the missing transverse momentum.

\begin{figure*}
\begin{center}
\begin{minipage}{1.7in}
\includegraphics[scale=0.8]{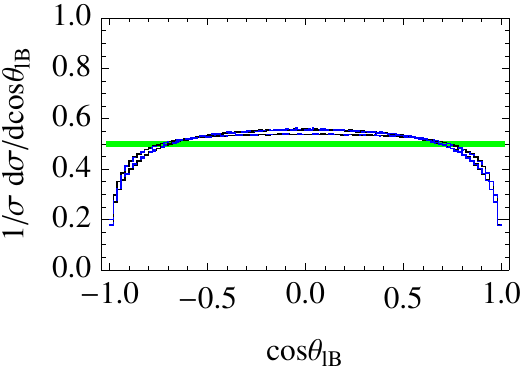}\\
(i)
\end{minipage}
\begin{minipage}{1.7in}
\includegraphics[scale=0.8]{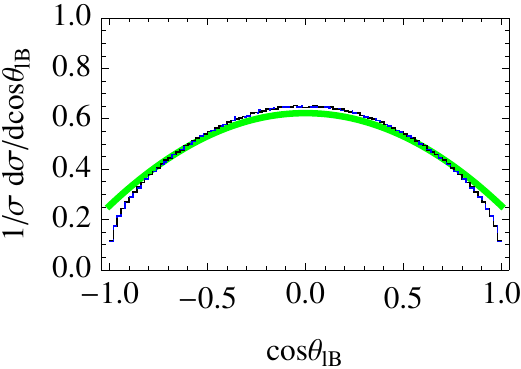}\\
(ii)
\end{minipage}
\begin{minipage}{1.7in}
\includegraphics[scale=0.8]{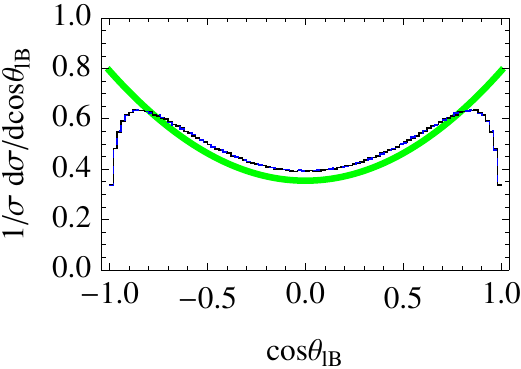}\\
(iiia)
\end{minipage}
\begin{minipage}{1.7in}
\includegraphics[scale=0.8]{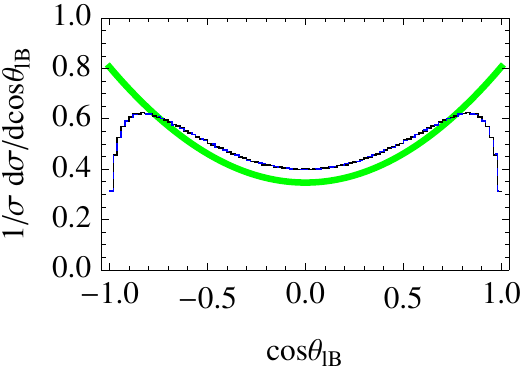}\\
(iiib)
\end{minipage}\\
\vspace{0.1in}
\begin{minipage}{1.7in}
\includegraphics[scale=0.8]{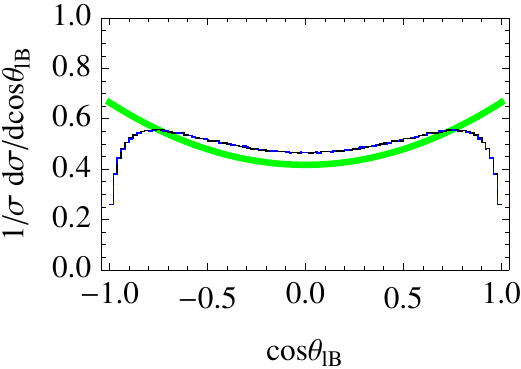}\\
(iiic)
\end{minipage}
\begin{minipage}{1.7in}
\includegraphics[scale=0.8]{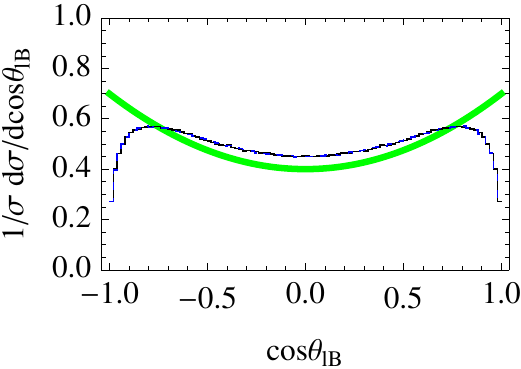}\\
(iiid)
\end{minipage}
\begin{minipage}{1.7in}
\includegraphics[scale=0.8]{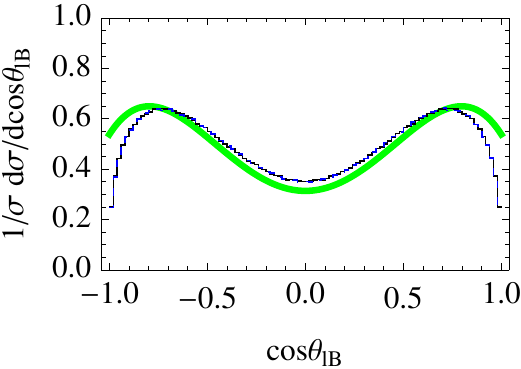}\\
(iva)
\end{minipage}
\begin{minipage}{1.7in}
\includegraphics[scale=0.8]{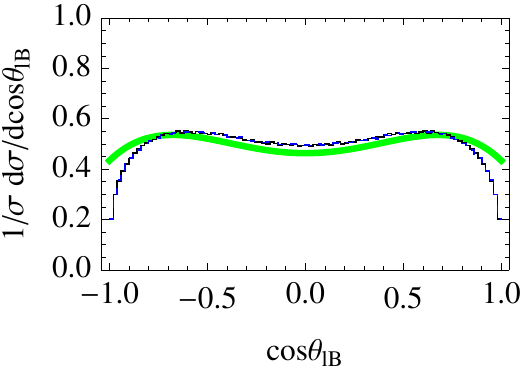}\\
(ivb)
\end{minipage}
\caption{\label{fig:ISR plots}The effect of ISR and beamstrahlung on the CM angular distributions.  The solid green lines come from the analytic formulas given in Table~\ref{tab:analytic CM angular distributions}, the solid black lines comes from the large solution given in Eq.~\eqref{eq:largesolution} where the ISR and beamstrahlung is turned on and the dashed blue lines, additionally, have the cuts in Eqs.~\eqref{eq:cos cut} through \eqref{eq:pT cut} applied.  The collision energy and masses are the same as in Fig.~\ref{fig:classes plots}.}
\end{center}
\end{figure*}

We display these effects in Fig.~\ref{fig:ISR plots}.
  The theoretical curves in the absence of ISR, beamstrahlung and cuts is shown in solid green.  The effect of including ISR and beamstrahlung is shown as the solid black lines and the additional effect of our basic cuts is shown by the dashed blue lines.  The solid black and dashed blue lines are reconstructed from the large solution, Eq.~\eqref{eq:largesolution}.  In all cases, the solid black lines and dashed blue lines coincide showing that these basic cuts do not have a significant impact on the CM angular distributions.  Of course, further cuts will be required to separate the signal from the background, but we leave that to a separate study.   ISR and beamstrahlung, on the other hand, do modify these distributions.  The modification is similar to that of the finite widths, but more extreme showing that the effect of finite widths is probably subdominant for measuring the CM angular distribution in this process.  The smearing due to ISR and beamstrahlung appears to be the dominant effect modifying the CM angular distributions from their theoretical form.  It reduces the edges and enhances the center of these distributions.
  
For the flat distributions (Fig.~\ref{fig:ISR plots}i), the edges are significantly reduced by approximately 60\% while the center is only enhanced by approximately 10\%.  In contrast to the effect of a finite width, the new distribution does not look parabolic and should be easily distinguishable.  For the concave negative parabola distribution (Fig.~\ref{fig:ISR plots}ii), the effect is small with a significant deviation from the theoretical curve only for $|\cos\theta_{\ell B}|\gtrsim0.9$.  For the concave positive parabola distributions (Fig.~\ref{fig:ISR plots}iii), the effect is to alter the distribution to look more like the ``M" distributions.  This could make distinguishing these two cases more difficult, however, it can be seen that the position of the peaks in the two cases is different as is the shape of the shoulder and the concavity of the central region.  With sufficient collision energy and luminosity, these cases should also be distinguishable.  For the theoretically ``M" shaped distribution (Fig.~\ref{fig:ISR plots}iv), the effect is to decrease the edges further while not changing much the central region.

\subsection{Comments}
In this section, we have illustrated the use of our kinematical variables for reconstructing the CM angular distribution in the process $e^+e^-\to\gamma^*/Z^*\to B^+B^-\to\mu^+\mu^-DD$.  We have found excellent agreement between our reconstruction and the true CM angular distribution, even in the presence of a small finite width.  Additionally, we showed that the curvature of the distributions depended on the collision energy, which is in principle controllable at the ILC.  We have further found that although ISR and beamstrahlung modify the distributions, this effect is predictable with Monte Carlo.  It appears as a reduction at the edges and a small enhancement in the center of the distribution.  For most of the classes, the essential shape can still be seen.  Therefore, in principle, it should be possible with sufficient luminosity to distinguish all the spin combinations using our kinematical variable except for those within Case i where $B$ is spin-0 or -$\frac{1}{2}$ or between spin combinations within the same class when the masses and collision energy conspire to make the distributions identical.  To end this section, we would like to discuss a few areas where the analysis of these kinematic variables applied to antler processes can be extended.

In this work, we have only considered parity symmetric S-channel production of the two $B$s.  In certain cases, parity violating operators may be important.  In these cases, we may be able to further separate the spin combination cases as well as get information about the parity violating nature of the couplings by using the small solution.  Furthermore, in some models, the T-channel may be important.  This is certainly true in the SM, where the process $e^+e^-\to W^+W^-\to\mu^+\nu_\mu\mu^-\bar{\nu}_\mu$ has an important contribution from a neutrino in the T-channel.  We have found that, in this SM process, the CM angular distribution when the T-channel diagram is included is sensitive to chiral couplings and that parity violation can be seen in the small solution.  A full analysis of the effect of parity violating operators and of the T-channel production will be carried out in a future publication.

We have assumed in this section that $D$ was self-charge-conjugate.  However, this does not have to be the case.  $D$ does not have to be its own antiparticle.  Although we have not done a full analysis of this possibility, we have found that the CM angular distribution in the case of the SM process $e^+e^-\to W^+W^-\to\mu^+\nu_\mu\mu^-\bar{\nu}_\mu$ is the same whether the neutrinos are Dirac or Majorana.

For our illustration, we used muons in the final state, however, our method would work just as well with jets.  The benefit of using muons is that they can be distinguished by their sign and their momentum can be measured more precisely than jets.  However, in some models, the branching ratio to jets could be much higher.  Although the jets are indistinguishable, it does not matter for the symmetric distributions we have presented here because each jet generates the same distribution.  In this case, the CM angular distribution coming from the large solution (Eq.~\eqref{eq:largesolution}) should be separately binned for each jet.  The final distribution will have the same shape as that for an individual jet but will be twice as high in each bin.

In this work, we have only considered the intermediate particle $B$ decaying to a two-body final state $L$ and $D$, however, it could also decay to multiple visible particles and $D$.  If the multiple visible particles come from another intermediate state that decays (such as a W decaying to jets) then there is no change to the results presented here.  The mass and spin component of $L$ are just those of the extra intermediate state (such as the W).  However, if, on the other hand, there is no intermediate state decaying to the multiple visible particles, $|\cos\theta_{LB}|$ can still be reconstructed for each event, but the CM angular distribution depends on the invariant mass $M_L$ which changes event to event.  It should, nevertheless, be possible to generate a series of CM angular distributions for each invariant mass range of the visible particles.  However, the situation is further complicated by the fact that the component of the angular momentum of the visible particles will not simply be the component of the spin of a single particle anymore, further complicating the interpretation of the CM angular distributions.


\section{\label{sec:conclusion}Conclusion}
In this work, we have constructed a new kinematical variable that gives, under certain circumstances, the CM angular distribution, $\cos\theta_{LB}$, of a decaying particle $B$ when one of its daughter particles $D$ is not detected.  Among the requirements for this method are that the masses of the mother particle $B$ as well as both daughter particles $L$ and $D$ must be known.  Additionally, we have shown that the magnitude of $\cos\theta_{LB}$ only depends on the component of the momentum of the measured particle $L$ which is transverse to the momentum of $B$, which we call $p_{\widetilde{T}}$, as can be seen in Eq.~\eqref{eq:cosmagfinal}.  We have also shown that $p_{\widetilde{T}}$ is uniquely related to the energy of the missed particle $D$, $E_D$, as seen in Eqs.~\eqref{eq:pTED1} and \eqref{eq:pTED2}.  Therefore, if either $p_{\widetilde{T}}$ or $E_D$ can be determined event by event, then the magnitude of $\cos\theta_{LB}$ is known, event by event, unambiguously.  

The sign of our observable, on the other hand, has an inherent two-fold ambiguity.  We find that it can be consistently split into two solutions which we call the large and small solutions, based on the size of the component of $\vec{p}_D$ which is parallel to $\vec{p}_L$.  We find that the large solution always has a negative sign, as in Eq.~\eqref{eq:largesolution} while the small solution's sign depends on the energy of the measured particle $L$, as in Eq.~\eqref{eq:smallsolution}.  Although this sign cannot be determined event by event, the distributions of the large and small CM angular distributions, Eq.~\eqref{eq:largesolution} and Eq.~\eqref{eq:smallsolution}, respectively, contain the information required to reconstruct the true $\cos\theta_{LB}$ in many situations.  The large solution gives the negative absolute value of the distribution of $\cos\theta_{LB}$, from which it is easy to reconstruct the full symmetrized version of the $\cos\theta_{LB}$ distribution by dividing it in half and taking the mirror image on the $\cos\theta_{LB}>0$ side.  Although this distribution can not tell us anything about the parity violation in the true $\cos\theta_{LB}$ distribution, it can give us the full symmetrized dependence of the differential cross-section on $\cos\theta_{LB}$ and therefore the spin-combinations of the mother and daughter particles which can produce that distribution.  In many cases, this is already sufficient to determine the spin of $B$ and $D$, where the spin of $L$ is already known.  Although the full structure of our kinematical variable was first described in the present work, a special case of this kinematical variable was applied to D-Y production of a charged lepton and a neutrino where the spin of the intermediate resonance was shown to be unambiguously determined by the large solution \cite{Chiang:2011kq}.

On the other hand, the small solution contains nontrivial information about the sign of $\cos\theta_{LB}$.  Although it does not agree with the true sign event by event, its distribution contains the clear signatures of the parity violation present in the true distribution.  In \cite{Chiang:2011kq}, it was shown that in the special case of D-Y production of a charged lepton and a neutrino, a simple reconstruction technique could be applied to the small solution to fully reconstruct the true CM angular distribution, including its parity violating features, almost exactly.  It was, further, shown that this reconstruction technique was universal and did not depend on the spin of $B$ or the parity-violation.  In other words, the reconstruction technique could be applied blindly and give the correct results in all spin cases.  Moreover, it was also shown that the parity violation could be determined directly from the small solution without applying the reconstruction technique.  

In \cite{Chiang:2011kq}, it was also shown that acceptance cuts, and even the rather large $p_T>250$GeV cut only affected the $|\cos\theta_{\ell R}|\gtrsim0.9$ edges of the distributions, but left the majority of the distribution unaffected, preserving its power.  In the present work, we extended this to consider the effect of a finite width.  We found that an approximately $1$\% width and smaller widths only affected the $|\cos\theta_{\ell R}|\gtrsim0.95$ edges and the $|\cos\theta_{\ell R}|<0.2$ central region, but left the rest of the distribution unchanged, as seen in Fig.~\ref{fig:DYlarge}.  This shows that this variable works quite well for realistic D-Y charged resonance production at the LHC.

In the present work, we applied our kinematical variables, for the first time, to the antler process $e^+e^-\to\gamma^*/Z^*\to B^+B^-\to\mu^+\mu^-DD$ at the ILC where $D$ was taken to be a self-charge-conjugate dark matter particle.  We showed that $E_D$ is known in this case, therefore $|\cos\theta_{LB}|$ is known, event by event, unambiguously, as described above.  
We focused, in this article, on only parity-symmetric operators, therefore all our angular distributions were inherently symmetric.
We calculated analytically the dependence of the differential cross-section on $\cos\theta_{LB}$ and included it in Table~\ref{tab:analytic CM angular distributions}.  We found agreement of the true CM angular distribution with these formulas in all cases.  The CM angular distributions split up into four classes.  The first class (denoted by i in Table~\ref{tab:analytic CM angular distributions} and Figs.~\ref{fig:classes plots} through \ref{fig:ISR plots})  is given by a flat distribution and, unfortunately, includes both the cases where $B$ is spin-$0$ and spin-$\frac{1}{2}$.  The second class (denoted by ii in Table~\ref{tab:analytic CM angular distributions} and Figs.~\ref{fig:classes plots} through \ref{fig:ISR plots}) is given by a concave negative parabola and includes only the case where $B$ is spin-$1$ and $D$ is spin-$\frac{1}{2}$.  The third class (denoted by iii in Table~\ref{tab:analytic CM angular distributions} and Figs.~\ref{fig:classes plots} through \ref{fig:ISR plots}) is given by a concave positive parabola and includes the case where $B$ is spin-$1$ and $D$ is spin-$\frac{3}{2}$ and the cases where $B$ is spin-$\frac{3}{2}$ and $D$ is spin-$0$, $1$ or $2$.  Although their distributions are all concave positive parabolas, we show that with appropriate masses and collision energy, they can be separated and distinguished as in Fig.~\ref{fig:iii splitting}iii.  The last class (denoted by iv in Table~\ref{tab:analytic CM angular distributions} and Figs.~\ref{fig:classes plots} through \ref{fig:ISR plots}) is given by a ``M" shape and includes the cases where $B$ is spin-$2$ and $D$ is spin-$\frac{1}{2}$ or $\frac{3}{2}$.  Again, we show that for appropriate parameters, these can be distinguished as in Fig.~\ref{fig:iii splitting}iv.

We then showed that the large CM angular distribution after a simple reconstruction, gives exact agreement with the true distribution in the narrow width limit and in the absence of ISR and beamstrahlung, as shown in Fig.~\ref{fig:classes plots}.  If the width is not infinitesimal, on the other hand, we found a small reduction on the edges and, to a lesser extent, a small enhancement in the center of the distribution occured.  However, for a $1$\% width, we found the effect to be practically negligible while a $5$\% width made a noticeable effect but left the shape of the distributions largely intact for most spin combinations, as seen in Fig.~\ref{fig:FW plots}.  The effect of ISR and beamstrahlung, on the other hand, was much more pronounced.  It also appeared as a reduction on the edges and, to a lesser extent, an enhancement in the center.  However, as can be seen in Fig.~\ref{fig:ISR plots}, the shape of the true distribution is still clearly visible for most of the spin combinations.  Furthermore, the effect of ISR and beamstrahlung can be well modeled.  From these results, we see that our method should work quite well in this process and that as long as $B$ is spin-$1$ or higher and the masses don't conspire to make the distributions within a class identical, it should be possible to determine the spin of both $B$ and the dark matter particle $D$ using our kinematical variables in this process at the ILC.

\textit{Acknowledgements} N.D.C. was supported in part by PITT PACC and the U.S. Department of Energy under grant No. DE-FG02-95ER40896.  D.S. was supported in part by PITT PACC, the Pennsylvania Space Grant Consortium Research Scholarship and the Dietrich School of Arts and Sciences Summer Undergraduate Research Award for Independent Research.  We would like to thank Tao Han, Adam Leibovich and Ayres Freitas for their encouragement, helpful discussions and support during the completion of this project.  We would also like to thank the University of Granada High Energy Theory Group for their hospitality during our visit where part of this research was completed.


\appendix

\section{\label{sec:Wigner d-functions}Wigner d-functions}
\begin{figure}
\begin{center}
\includegraphics[scale=0.4]{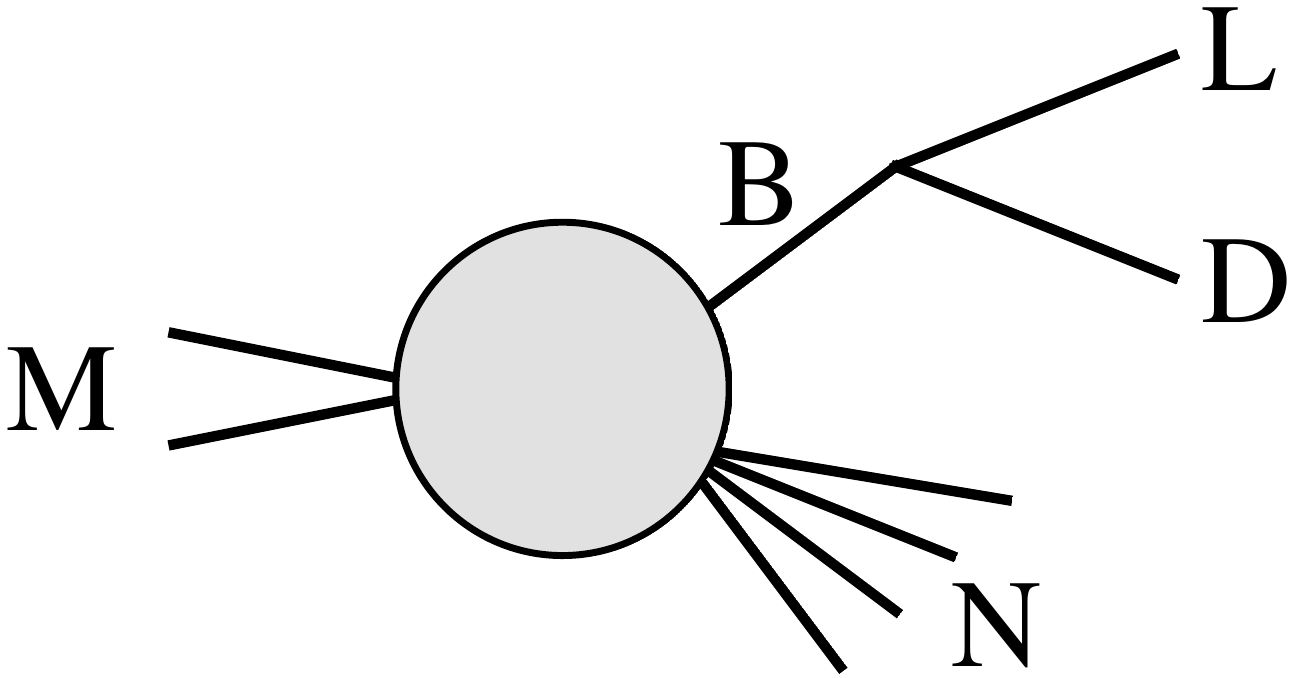}
\end{center}
\caption{\label{fig:Wigner diagram}Diagram of a scattering process dominated by an on-shell $B$ which decays to $L$ and $D$.}
\end{figure}
In this appendix, we give a very rough description of the dependence of the differential cross section on the Wigner d-functions.
Consider a scattering process that is dominated by diagrams where an intermediate particle $B$ is on-shell and decays to $L$ and $D$ as in Fig.~\ref{fig:Wigner diagram}.  Since $B$ is on-shell, its propagator numerator is equal to a sum over the spin-z components of dyads of its polarization vector, as in
\begin{equation}
\lim_{p_B^2\to M_B^2}\Pi\left(p_B^2\right) = \sum_{\sigma_B}\epsilon_{\sigma_B}\left(p_B\right)\epsilon_{\sigma_B}^*\left(p_B\right)
\end{equation}
where $\Pi\left(p_B^2\right)$ is $B$'s propagator numerator and $\epsilon_{\sigma_B}\left(p_B\right)$ is $B$'s polarization vector for spin-z component $\sigma_B$.  
(We use $\epsilon_i\left(p\right)$ here to represent the polarization vector or spinor for any spin.)
Using this, the amplitude can be written as
\begin{equation}
\mathcal{M}_{\sigma_M\sigma_N\sigma_L\sigma_D} = \sum_{\sigma_B}\widetilde{\mathcal{M}}_{\sigma_M\sigma_N\sigma_B}\epsilon_{\sigma_B}^*\left(p_B\right)\Gamma\ \epsilon_{\sigma_L}\left(p_L\right)\epsilon_{\sigma_D}\left(p_D\right)
\end{equation}
where $\widetilde{\mathcal{M}}$ contains everything not explicitly written in this equation, including the polarization vectors of all the other external states, the polarization vector for $B$ and the propagator denominator for $B$.  $\Gamma$ represents the vertex factor for the $BLD$ system and may connect to the polarization vectors and also may connect to $\widetilde{\mathcal{M}}$.  

We will consider the $BLD$ system in the $B$ CM frame.  We take the $z$-direction to be along $\vec{p}_B$ and take $\sigma_B$ to be the component of $B$'s spin along this direction.  Since this system consists solely of $B$ before the decay, the total angular momentum is equal to the spin of $B$.  Because total angular momentum is conserved, the final state $LD$ must have the same total angular momentum.  If we measure the final total angular momentum component along $\vec{p}_L$, it will be equal to the difference of the helicities of $L$ and $D$, which we will call $\sigma_L$ and $\sigma_D$, respectively.  Although orbital angular momentum in the $LD$ system can contribute to the total angular momentum, its component is perpendicular to $\vec{p}_L$ and, therefore, will not contribute to the component of the total angular momentum along that direction.  Since total angular momentum is conserved, the dependence on this angle is simply the quantum mechanical overlap between these bases,
\begin{eqnarray}
d^{s_B}_{\sigma_B,\sigma_L-\sigma_D}\left(\theta_{LB}\right) &=& \langle s_B,\sigma_L-\sigma_D,\theta_{LB} | s_B,\sigma_B \rangle\\
&=& \langle s_B,\sigma_L-\sigma_D | e^{iJ_\perp \theta_{LB}} | s_B,\sigma_B \rangle\nonumber
\end{eqnarray}
where $J_\perp$ is the component of the angular momentum operator perpendicular to the $BLD$ system.  These functions are called the Wigner d-functions.  With this, in the $B$ CM frame, we have
\begin{equation}
\epsilon_{\sigma_B}^*\left(p_B\right)\Gamma\ \epsilon_{\sigma_L}\left(p_L\right)\epsilon_{\sigma_D}\left(p_D\right) \propto d^{s_B}_{\sigma_B,\sigma_L-\sigma_D}\left(\theta_{LB}\right)
\end{equation}

Putting these things together gives
\begin{equation}
\mathcal{M}_{\sigma_M\sigma_N\sigma_L\sigma_D} = \sum_{\sigma_B}\widetilde{\widetilde{\mathcal{M}}}_{\substack{\sigma_M\sigma_N\sigma_B\\\sigma_L\sigma_D}}d^{s_B}_{\sigma_B,\sigma_L-\sigma_D}\left(\theta_{LB}\right)
\end{equation}
where $\widetilde{\widetilde{\mathcal{M}}}$ absorbs all other factors.
After squaring, averaging over initial spins, summing over final spins, multiplying by phase factors and integrating over all other momentum factors, we have
\begin{equation}
\frac{d\sigma}{d\theta_{LB}} = \sum_{{\substack{\sigma_B\sigma'_B\\\sigma_L\sigma_D}}} \mathcal{A}_{\substack{\sigma_B\sigma'_B\\\sigma_L\sigma_D}}
d^{s_B}_{\sigma_B,\sigma_L-\sigma_D}\left(\theta_{LB}\right)d^{s_B}_{\sigma'_B,\sigma_L-\sigma_D}\left(\theta_{LB}\right)
\end{equation}
We see that the dependence on this angle is restricted by conservation of angular momentum to be proportional to a sum of squares of the Wigner $d^j_{m'm}$-functions, where $j=s_B$, $m=\sigma_B$ or $\sigma'_B$ and $m'=\sigma_L-\sigma_D$, independent of the model.  The coefficients $\mathcal{A}_{\sigma_B\sigma'_B\sigma_L\sigma_D}$, on the other hand, are dependent on the couplings of the model as well as the polarization of the beams.

\section{\label{sec:pdz}Derivation of $p_{Dz}$}

In this appendix, we derive $p_{Dz}$, the z-component of the $D$ momentum.  Since the $z$-axis was defined to be in the same direction as the momentum of $B$, the net momentum transverse to this direction must be zero.  Without loss of generality, we will take $p_{\widetilde{T}}$ to be along the $x$-axis.  We can then define the momentum of $L$ and $D$ as
\begin{eqnarray}
p_L &=& \left( E_L, p_{\widetilde{T}},0,p_{Lz} \right)\ , \\
p_{D} &=& \left( E_D, -p_{\widetilde{T}},0,p_{Dz} \right)\ .
\end{eqnarray}
The invariant mass of $B$ is then given by
\begin{eqnarray}
M_B^2 &=& M_L^2+M_D^2+2E_L\sqrt{M_D^2+p_{\widetilde{T}}^2+p_{Dz}^2}\nonumber\\
&&+2p_{\widetilde{T}}^2-2p_{Lz}p_{Dz}\ .
\end{eqnarray}
This equation can be solved for $p_{Dz}$ giving
\begin{equation}
p_{Dz} = p_{Lz}\left( \frac{M_B^2+M_L^2-M_D^2}{2\left(M_L^2+p^2_{\widetilde{T}}\right)}-1\right) \pm \frac{E_L\Delta}{2\left(M_L^2+p^2_{\widetilde{T}}\right)}\ ,
\end{equation}
where
\begin{equation}
\Delta = \left(M_B^2+M_L^2-M_D^2\right)\sqrt{1-\frac{4M_B^2\left(M_L^2+p^2_{\widetilde{T}}\right)}{\left(M_B^2+M_L^2-M_D^2\right)^2}}\ .
\end{equation}
Recalling the relation for $E_{LCM}$ (Eq. \eqref{eq:elcms}), we have
\begin{eqnarray}
p_{Dz}&=&p_{Lz}\left(\frac{M_BE_{LCM}}{M_L^2+p^2_{\widetilde{T}}}-1\right)\nonumber \\
&&\pm \frac{M_BE_LE_{LCM}}{M_L^2+p^2_{\widetilde{T}}}\sqrt{1-\frac{M_L^2+p^2_{\widetilde{T}}}{E^2_{LCM}}}\ .
\end{eqnarray}
We note that our result does not depend on the choice of $x$- and $y$-axis.


\section{\label{sec:EL>ELCM}Relationship of \boldmath{$E_L$} and \boldmath{$E_{LCM}$} for the small solution}
In this appendix, we assume the small solution.   Using Eq.~\eqref{eq:pbz} when $p_{Lz}>0$, we have 
\begin{equation}
\sign{p_{Bz}}_{>}=\sign{p_{Lz}^2-E_L^2\zeta^2}_>>0\ ,
\end{equation}
for the small solution.
Plugging in $\zeta$ (see Eq.~\eqref{eq:sigma}), using $E_L^2-p_{Lz}^2=M_L^2+p_{\widetilde{T}}^2$ and factoring out the positive $\left(M_L^2+p_{\widetilde{T}}^2\right)$ gives
\begin{equation}
\sign{\frac{E_L^2}{E^2_{LCM}}-1}_>>0\label{eq:EL>ELCM}\ .
\end{equation}

On the other hand, when $p_{Lz}<0$, we have
\begin{equation}
\sign{p_{Lz}+E_L\zeta}_<>0\ ,
\end{equation}
for the small solution, which can be rewritten as
\begin{equation}
\sign{-p_{Lz}^2+E_L^2\zeta^2}_<>0\ .
\end{equation}
Following the same manipulations as above brings this to the form
\begin{equation}
\sign{-\frac{E_L^2}{E^2_{LCM}}+1}_<>0\label{eq:EL<ELCM}\ ,
\end{equation}
or
\begin{equation}
\sign{\frac{E_L^2}{E^2_{LCM}}-1}_<<0\label{eq:EL<ELCM2}\ .
\end{equation}

From these two results, we learn that
\begin{equation}
\sign{p_{Lz}} = \sign{E_L-E_{LCM}}\ .
\end{equation}


\section{\label{sec:twofold ambiguity}Two-Fold Ambiguity of Antler Momenta at the ILC}
In this section, we consider processes at the ILC of the type shown in Fig.~\ref{fig:antler}, dubbed antler diagrams.  Following \cite{Han:2009ss,Christensen:2014yya}, we will assume that the masses of $B$ and $D$ are known and show that there is only a discrete ambiguity in the momenta of the two $D$s.  We begin by noting that there are eight unknowns in the momenta of the two $D$s and count the constraints.  Due to conservation of energy and the symmetry of the diagram, we find:
\begin{equation}
\frac{\sqrt{s}}{2} = E_{B_i} = E_{l_i}+E_{D_i}\ ,
\end{equation}
where $i$ refers to which $B$ and $l$ in Fig.~\ref{fig:antler}.  This implies the two linear equations
\begin{equation}
E_{D_i} = \frac{\sqrt{s}}{2}-E_{l_i}\label{ED1CMS}\ ,
\end{equation}
and completely specifies the energy of the two $D$s.

We also have the conservation of three-momenta which gives the three linear equations
\begin{equation}
\vec{p}_{D_1}+\vec{p}_{D_2} = -\vec{p}_{l_1}-\vec{p}_{l_2}\label{CMS p cons}\ .
\end{equation}

For the mass of B, we have $\left(p_D+p_l\right)^2$ which reduces to
\begin{equation}
M_B^2 = M_D^2 + 2E_{l_i}E_{D_i} - 2\vec{p}_{l_i}\cdot\vec{p}_{D_i}\ .
\end{equation}
After plugging in Eq.~\eqref{ED1CMS}, this gives us the two following linear equations
\begin{equation}
2\vec{p}_{l_i}\cdot\vec{p}_{D_i} = M_D^2-M_B^2+\sqrt{s}E_{l_i}-2E_{l_i}^2\label{eq:pl1.pD1}\ .
\end{equation}

Finally, we have the mass of D which, after substituting Eq.~\eqref{ED1CMS}, gives the two quadratic equations
\begin{equation}
\left|\vec{p}_{D_i}\right|^2 = \left(\frac{\sqrt{s}}{2}-E_{l_i}\right)^2-M_D^2\ .
\end{equation}

All together this makes nine equations for eight unknowns.  Seven are linear and two are quadratic, one of which is linearly dependent.  This leaves us with two discrete solutions.  For further details, see \cite{Tsukamoto:1993gt,Choi:2006mr}.

\section{\label{sec:effective operators}Effective Operators}

In this appendix we provide a list of all the parity-conserving operators used in our analysis, except those already described in the SM. The naming conventions for the fields are given in Table \ref{table:particlenames}.  
As described in Sec.~\ref{sec:antlers}, we included the majority of parity-conserving operators in our analysis in a model independent way.  There were only a few parity-conserving operators that we did not include and we specify them as they come up in this appendix.  These operators will be considered in a future publication.

\subsection{Interactions of $\mathbf{B}$ with the Photon}

These operators follow directly from QED, where we replace the partial derivative in the kinetic term with the covariant derivative, denoted here by $\mathcal{D}$, to ensure that our Lagrangians are manifestly QED invariant. For spin-0, we have
\begin{equation}
\mathcal{L}_{\sB} = -\sB^*\left(\mathcal{D}^2+m^2\right)\sB.
\end{equation}
For spin-$\frac{1}{2}$, we have
\begin{equation}
\mathcal{L}_{\fB} = \overline{\fB}\left(i\slashed{\mathcal{D}}-m\right)\fB.
\end{equation}
For spin-$1$, we have
\begin{equation}
\mathcal{L}_{\vB} = \vB^{\mu *}\left(\mathcal{D}^2{\eta}_{\mu \nu}-\mathcal{D}_{\mu}\mathcal{D}_{\nu}+m^2\eta_{\mu \nu}\right)\vB^{\nu}.
\end{equation}

The Lagrangian for a spin-3/2 field was first reported by Rarita and Schwinger \cite{Rarita:1941mf}. However, it is now customary to write the Lagrangian in a way that is more compact, although entirely equivalent. In our work, we use the Lagrangian given in \cite{Weinberg:2000cr},
\begin{equation}
\mathcal{L}_{\rB} = \overline{\rB}_{\mu}\left(\epsilon ^{\mu \alpha \beta \nu}{\gamma}_{\alpha}{\gamma}_5\mathcal{D}_{\beta}-m\left({\eta}^{\mu \nu}-{\gamma}^{\mu}{\gamma}^{\nu}\right)\right)\rB_{\nu}.
\end{equation}

For spin-$2$, we have \cite{Barua:1978ck}
\begin{equation}
\mathcal{L}_{\tB} = \tB^{\mu \nu *}\Phi_{\mu \nu \rho \sigma}\tB^{\rho \sigma}\ ,
\end{equation}
where
\begin{align}
6\Phi^{\mu \nu \rho \sigma} \equiv& \mathcal{D}^2{\eta}_{\mu \nu}{\eta}_{\rho \sigma}-\mathcal{D}^2{\eta}_{\mu \sigma}{\eta}_{\nu \rho}-{\eta}_{\mu \nu}\mathcal{D}_{\rho}\mathcal{D}_{\sigma}\nonumber\\&+{\eta}_{\mu \sigma}\mathcal{D}_{\nu}\mathcal{D}_{\rho}+{\eta}_{\nu \sigma}\mathcal{D}_{\mu}\mathcal{D}_{\rho}-{\eta}_{\rho \sigma}\mathcal{D}_{\mu}\mathcal{D}_{\nu}\nonumber\\&+m^2\left({\eta}_{\mu \nu}{\eta}_{\rho \sigma}-{\eta}_{\mu \sigma}{\eta}_{\nu \rho}\right)\ .
\end{align}
We could have also included a term of the form $g_{_{A\tB}} F^{\mu \nu}{\tB_{\mu}}^{\rho *}\tB_{\nu \rho}$ which is also parity conserving, however we find that this operator gives a more complicated form of the symmetric CM angular distribution.  Specifically, $\mathcal{A}_{iv}$, $\mathcal{B}_{iv}$, $\mathcal{C}_{iv}$ and $\mathcal{N}_{iv}$ from Table~\ref{tab:analytic CM angular distributions} depend on the value of $g_{_{A\tB}}$ if non-zero.  For simplicity, in the current article, we choose to set $g_{_{A\tB}}=0$ so that $\mathcal{A}_{iv}$, $\mathcal{B}_{iv}$, $\mathcal{C}_{iv}$ and $\mathcal{N}_{iv}$ depend only on the masses and collision energy.

\subsection{Interactions of $\mathbf{B}$ with the $\mathbf{Z}$-boson}
For convenience, we implemented the photon interactions of the previous subsection with the hypercharge gauge boson.  Thus, the covariant derivative is of the form
\begin{equation}
\mathcal{D}_{\mu}=\partial_\mu+i g' B_\mu=\partial_{\mu}+ieA_{\mu}-i g'\sin\theta_W Z_\mu\ .
\end{equation}
Therefore, the operators of the previous subsection induce interactions with the Z boson as well.  In addition to these operators, we also included the Lagrangian terms as described below.

For spin-$0$, we include the Lagrangian 
\begin{equation}
\mathcal{L}_{\sB}=g_{_{Z\sB}} |\sB|^2 {\partial_{\mu}} Z^{\mu}\ .
\end{equation}
For spin-$\frac{1}{2}$, we have 
\begin{equation}
\mathcal{L}_{\fB}=g_{_{Z\fB}} Z^{\mu} \overline{fB} {\gamma _{\mu}} \fB\ .
\end{equation}
For spin-$1$, we include
\begin{equation}
\mathcal{L}_{\vB}= g_{_{Z\vB}} {\vB}_{\mu}^* {\vB}^{\mu} {\partial}_{\nu} Z^{\nu} + \hc\ .
\end{equation}
We could have also included the operator $g_{_{Z\vB 2}} {\vB}^{\mu *} {\vB}^{\nu} {\partial}_{\mu} Z_{\nu} + \hc$ which is also parity conserving, however we find that, as for the case where $B$ is spin-2 coupling to a photon, this operator gives a more complicated form of the symmetric CM angular distribution.  Specifically, $\mathcal{A}_{ii}$, $\mathcal{B}_{ii}$, $\mathcal{N}_{ii}$, $\mathcal{A}_{iiia}$, $\mathcal{B}_{iiia}$ and $\mathcal{N}_{iiia}$ from Table~\ref{tab:analytic CM angular distributions} depend on the value of $g_{_{Z\vB 2}}$ if non-zero.  For simplicity, in the current article, we choose to set $g_{_{Z\vB 2}}=0$ so that $\mathcal{A}_{ii}$, $\mathcal{B}_{ii}$, $\mathcal{N}_{ii}$, $\mathcal{A}_{iiia}$, $\mathcal{B}_{iiia}$ and $\mathcal{N}_{iiia}$ depend only on the masses and collision energy.

For spin-$\frac{3}{2}$, we include the operator
\begin{equation}
\mathcal{L}_{\rB} = g_{_{Z\rB}} Z^{\mu} \overline{\rB}^{\nu} {\gamma}_{\mu} {\rB}_{\nu}\ .
\end{equation}
For spin-$2$, we include
\begin{equation}
\mathcal{L}_{\tB}= g_{_{Z\tB 1}} {\tB}^{\mu \nu *} {\tB}_{\mu \nu} {\partial}^{\lambda} Z_{\lambda} + g_{_{Z\tB 2}} {\partial}^{\lambda} {\tB}^{\mu \nu *} {\tB}_{\mu \nu} Z_{\lambda}   + \hc\ .
\end{equation}
We could have also included the operators $g_{_{Z\tB 3}} {\tB}^{\mu \nu *} {\tB}_{\mu \lambda} {\partial}^{\lambda} Z_{\nu} + g_{_{Z\tB 4}} {\partial}^{\mu} {\tB}^{\nu \lambda *} {\tB}_{\mu \nu} Z_{\lambda} + \hc$ which are also parity conserving, however we find that, as for the case where $B$ is spin-1, these operators give a more complicated form of the symmetric CM angular distribution.  Specifically, $\mathcal{A}_{iv}$, $\mathcal{B}_{iv}$, $\mathcal{C}_{iv}$ and $\mathcal{N}_{iv}$ from Table~\ref{tab:analytic CM angular distributions} depend on the values of $g_{_{Z\tB 3}}$ and $g_{_{Z\tB 4}}$ if non-zero.  For simplicity, in the current article, we choose to set $g_{_{Z\tB 3}}=g_{_{Z\tB 4}}=0$ so that $\mathcal{A}_{iv}$, $\mathcal{B}_{iv}$, $\mathcal{C}_{iv}$ and $\mathcal{N}_{iv}$ depend only on the masses and collision energy.

\subsection{Interactions of $\mathbf{B}$ with $\mathbf{\ell}$ and $\mathbf{D}$}

We begin with the Lagrangian for the case where $B$ is spin-$0$,
\begin{align}
\mathcal{L}_{\sB}=&g_{_{\sB\ell \fD}} \sB\ \overline{\ell} \fD  + g_{_{\sB\ell \rD}} \sB\ \overline{\ell} {\gamma}^{\mu} \rD_{\mu} + \hc\ .
\end{align}

For the case in which $B$ is spin-$\frac{1}{2}$, the Lagrangian reads
\begin{align}
\mathcal{L}_{\fB} =& g_{_{\fB\ell \sD}} \sD\ \overline{\ell} \fB + g_{_{\fB\ell \vD}} {\vD}^{\mu}\ \overline{\ell} {\gamma}_{\mu} \fB \nonumber\\
				  & + g_{_{\fB\ell \tD}} {\tD}^{\mu \nu} {\partial}_{\mu} \overline{\ell} {\gamma}_{\nu} \fB + \hc\ .
\end{align}

In the case where $B$ is spin-$1$, we have
\begin{align}
\mathcal{L}_{\vB} =& g_{_{\vB\ell \fD}} {\vB}^{\mu}\ \overline{\ell} {\gamma}_{\mu} \fD + g_{_{\vB\ell \rD1}} {\vB}^{\mu}\ \overline{\ell} {\rD}_{\mu} \nonumber\\
				  & + g_{_{\vB\ell \rD2}} {\vB}^{\mu}\ \overline{\ell} {\sigma}_{\mu \nu} {\rD}^{\nu} + \hc\ .
\end{align}

For a spin-$\frac{3}{2}$ $B$, we have
\begin{eqnarray}
\mathcal{L}_{\rB} &=& g_{_{\rB\ell \sD}} \sD\ {\partial}_{\mu} \overline{\ell} {\rB}^{\mu} + g_{_{\rB\ell \vD}} {\vD}^{\mu}\ \overline{\ell} {\rB}_{\mu} \nonumber\\
				 & & + g_{_{\rB\ell \tD}} {\tD}^{\mu \nu}\ \overline{\ell} {\gamma}_{\mu} {\rB}_{\nu} + \hc\ .
\end{eqnarray}

Finally, we have for the spin-$2$ particle
\begin{equation}
\mathcal{L}_{\tB} = g_{_{\tB\ell \fD}} {\tB}^{\mu \nu} {\partial}_{\mu} \overline{\ell} {\gamma}_{\nu} \fD + g_{_{\tB\ell \rD}} {\tB}^{\mu \nu} \overline{\ell} {\gamma}_{\mu} {\rD}_{\nu} +\hc\ .
\end{equation}

\section{\label{sec:CM analytic derivation}True On-Shell CM Angular Distribution Formulas}
\begin{figure}
\includegraphics[scale=.5]{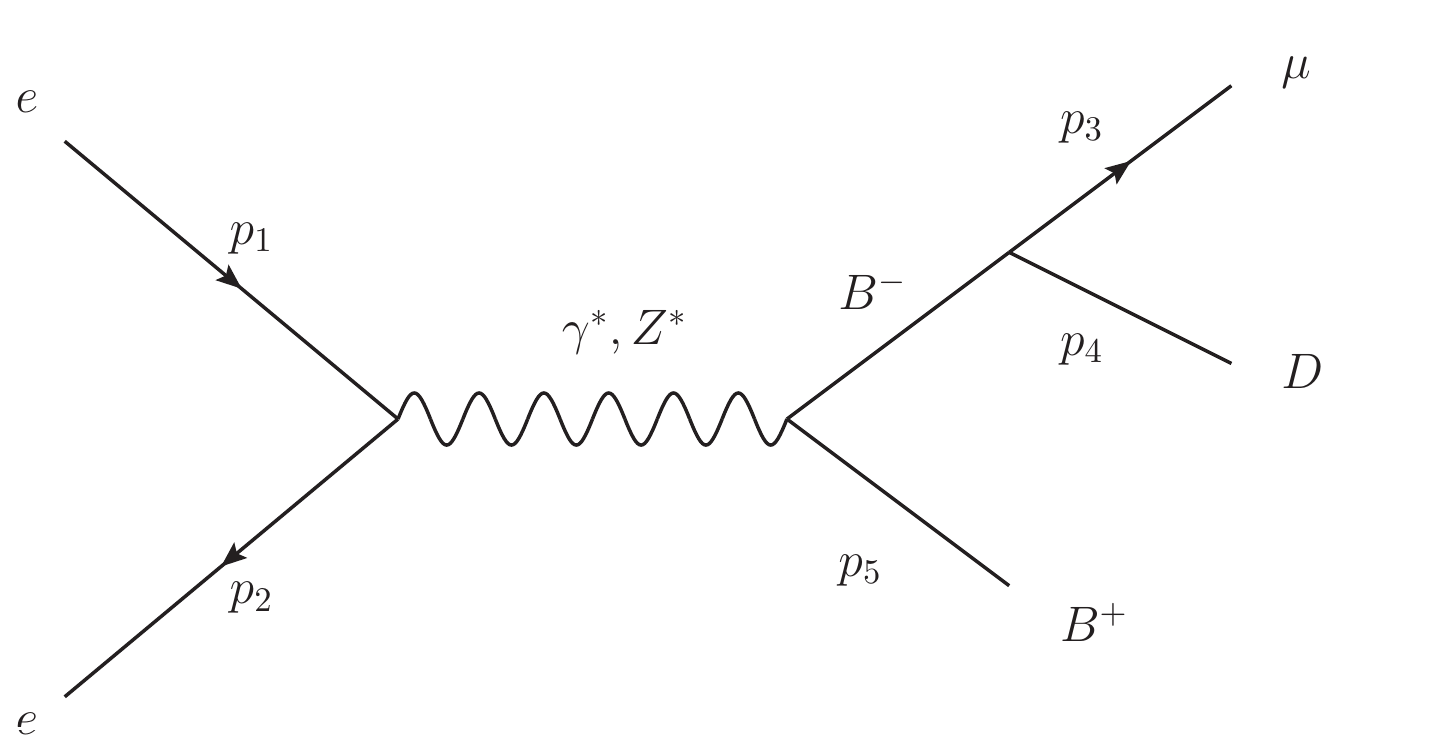}
\caption{\label{fig:2->3 antler}Pair production of $B^+$ and $B^-$ followed by the decay of $B^-$ to $\mu^-D$. }
\end{figure}

In this section, we describe the derivation of the formulas in Table~\ref{tab:analytic CM angular distributions}.  Using the CalcHEP implementation of our new particles $B$ and $D$ along with their interactions, we generated the analytic expressions for the squared amplitude of the diagram shown in Fig.~\ref{fig:2->3 antler}.
From there, we exported these expressions to Mathematica.  (In the CalcHEP symbolic interface, after squaring the diagrams choose ``Symbolic calculations" followed by ``Mathematica code".  The resulting Mathematica files will be in the results directory \cite{Belyaev:2012qa}.)  At this point, we had the squared amplitude in terms of masses, collision energy and inner products of the external momenta.  We note that we included the photon-photon squared diagram, the Z-Z squared diagram and the interference photon-Z and Z-photon squared diagrams in our calculation.  We removed the Z width from the calculation since its momentum was far off-shell, however we included the width of the $B$ in the calculation, which was necessary since we assume $B$ is on-shell for these formulas.    By conservation of momenta, $p_5$ can be removed from these expressions so we only need to deal with $p_1$ through $p_4$.  

Since each of the inner products is an invariant, we can do this calculation in any reference frame.  We choose the $z$-direction to be along the direction of $\vec{p}_B=\vec{p}_3+\vec{p}_4$ so that in the lab frame, $p_B=(E_B,0,0,|\vec{p}_B|)$.  We then boost into the $B$ CM frame.  In this frame, 
\begin{eqnarray}
p_{BCM} &=& M_B\left(1,0,0,0\right)\ ,\\
p_{3CM} &=& E_{LCM}\left(\begin{array}{c}1\\\sin\theta_{LB}\cos\phi_{LB}\\\sin\theta_{LB}\sin\phi_{LB}\\\cos\theta_{LB}\end{array}\right)\ ,\\
p_{4CM} &=& E_{LCM}\left(\begin{array}{c}\frac{E_{DCM}}{E_{LCM}}\\-\sin\theta_{LB}\cos\phi_{LB}\\-\sin\theta_{LB}\sin\phi_{LB}\\-\cos\theta_{LB}\end{array}\right)\ ,
\end{eqnarray}
where, by the usual conservation of momentum arguments, 
\begin{eqnarray}
E_{LCM} &=& \frac{M_B^2-M_D^2}{2M_B}\ ,\\
E_{DCM} &=& \frac{M_B^2+M_D^2}{2M_B}\ .
\end{eqnarray}

The initial momenta are known in the lab frame.  They are
\begin{eqnarray}
p_1 &=& \frac{\sqrt{s}}{2}\left(\begin{array}{c}1\\\sin\theta_{12}\cos\phi_{12}\\\sin\theta_{12}\sin\phi_{12}\\\cos\theta_{12}\end{array}\right)\ ,\\
p_2 &=& \frac{\sqrt{s}}{2}\left(\begin{array}{c}1\\-\sin\theta_{12}\cos\phi_{12}\\-\sin\theta_{12}\sin\phi_{12}\\-\cos\theta_{12}\end{array}\right)\ ,
\end{eqnarray}
where the angle $\theta_{12}$ is with respect to the direction of $\vec{p}_B$.  After boosting into the $B$ CM frame, we have
\begin{eqnarray}
p_{1CM} &=& \frac{\sqrt{s}}{2}\left(\begin{array}{c}
\frac{E_B}{M_B}-\frac{|\vec{p}_B|}{M_B}\cos\theta_{12}\\
\sin\theta_{12}\cos\phi_{12}\\
\sin\theta_{12}\sin\phi_{12}\\
\frac{E_B}{M_B}\cos\theta_{12}-\frac{|\vec{p}_B|}{M_B}
\end{array}\right)\ ,\\
p_{2CM} &=& \frac{\sqrt{s}}{2}\left(\begin{array}{c}
\frac{E_B}{M_B}+\frac{|\vec{p}_B|}{M_B}\cos\theta_{12}\\
-\sin\theta_{12}\cos\phi_{12}\\
-\sin\theta_{12}\sin\phi_{12}\\
-\frac{E_B}{M_B}\cos\theta_{12}-\frac{|\vec{p}_B|}{M_B}
\end{array}\right)\ ,
\end{eqnarray}
where $|\vec{p}_B|=\sqrt{E_B^2-M_B^2}$ and $E_B=\sqrt{s}/2$.  We plugged these expressions for the momenta into the squared amplitude formulas.  Since the differential cross section is equal to this squared amplitude times factors that are independent of these angles, we next integrated our expression over the angles $\theta_{12},\phi_{12}$ and $\phi_{LB}$.  This left us with an expression in terms of only $M_B,M_D,\sqrt{s}$ and $\theta_{LB}$.  We finally normalized this expression by dividing by the same expression integrated over $\theta_{LB}$.  This gave us the normalized differential cross-sections found in Table~\ref{tab:analytic CM angular distributions}.


\end{document}